\documentclass[%
 reprint,
 amsmath,amssymb,
 aps,
]{revtex4-2}

\usepackage{graphicx}% Include figure files
\usepackage{dcolumn}% Align table columns on decimal point
\usepackage{bm}% bold math
\usepackage{hyperref}% add hypertext capabilities
\usepackage{float}
\usepackage{comment}
\begin{document}

\preprint{APS/123-QED}

\title{Electron Vortex Beams for Chirality Probing at the Nanoscale}

\author{Neli Laštovičková Streshkova}
\email[]{neli.streshkova@matfyz.cuni.cz}
\author{Petr Koutenský}
\author{Martin Kozák}
\affiliation{Department of Chemical
Physics and Optics, Faculty of Mathematics and Physics, Charles University, Ke Karlovu 3, Prague CZ-12116, Czech Republic.}%

\date{\today}% It is always \today, today,

\begin{abstract}
 In this work we propose a new method for probing the chirality of nanoscale electromagnetic near-fields utilizing the properties of a coherent superposition of free electron vortex states in electron microscopes. Electron beams optically modulated into vortices carry orbital angular momentum (OAM), thanks to which they are sensitive to the spatial phase distribution and topology of the investigated field. The sense of chirality of the studied specimen can be extracted from the spectra of the electron beam with nanoscale precision owing to the short picometer de Broglie wavelength of the electron beam. We present a detailed case study of the interaction of a coherent superposition of electron vortex states and the optical near-field of a golden nanosphere illuminated by circularly polarized light as an example and we examine the chirality sensitivity of EVBs on intrinsically chiral plasmonic nanoanteannae. 
\end{abstract}

\maketitle

%\tableofcontents

\section{\label{sec1:level1}Introduction}

Chirality plays an important role in nature motivating the search for reliable and robust methods that allow the measurement of chiral phenomena on the nanoscale. Most routinely they are explored by optical techniques of the family of circular dichroism (CD), where the difference between the optical response of the studied system to right-handed and left-handed circularly polarized light is measured. In bio-chemistry and pharmacy, ultraviolet circular dichroism (UV-CD) is used to study the secondary structure of proteins, like $\alpha$-helixes and $\beta$-sheets \cite{joh1990}. Ultrafast time-resolved approaches are being developed to study chiral molecules with enhanced sensitivity such as photoelectron circular dichroism (PECD) \cite{bow2001}, photoexcitation circular dichroism (PXCD) \cite{bea2018} and enantio-selective nonlinear optics \cite{Ayuso2021, Ayuso2022}. In solid-state physics, X-ray CD is used to study the magnetic structure of materials \cite{sto1998}. Metamaterials and plasmonic structures with tailored geometries are studied by UV-CD \cite{sar2019}. However, as these techniques rely on optical probes, they are limited in resolution by the scales of the wavelength of the probing field, thus only allowing for the measurement of ensembles of chiral structures or chiral molecules and aggregates.
By using an electron wave it is in principle possible to achieve sub-nanometer resolution and to study the chirality of single nanoobjects like molecules or defects in thin crystals.

Besides electrostatic \cite{ver2018,poz2020} and magnetostatic \cite{beh2014} elements and holographic masks \cite{blo2013, gri2014}  used to shape the electron beam profile, recently great interest has been directed towards electron wave shaping via coherent interaction with light \cite{sch2019, fei2015, van2018, kon2020, fei2020, aba2021, chi2022}. With the variety of available tools, we can readily shape the electron beams to give them the desired properties for chirality probing.

The chirality of an object manifests once it interacts with another chiral object. While for light waves, circularly polarized light appears to be the most obvious example of a chiral probe, in the case of free electrons, vortex states are a promising candidate \cite{ver2010, BLIOKH20171}. Vortex states in general are associated with a phase factor $e^{il\varphi}$, with $\varphi$ being the azimuthal angle in cylindrical coordinates and $l$ being the so-called topological charge of the vortex. In this article we refer to the coherent superposition of such vortex states with different momenta and OAM as an electron vortex beam. Electron vortices can be readily generated in scanning and transmission electron microscopes using magnetic monopoles \cite{beh2014}, phase plates \cite{uch2010, tav2022} and holographic masks \cite{ver2010, yu2023}. Interaction with light fields allows for ultrafast generation and control of electron vortices in the vicinity of plasmonic nanostructures \cite{van2019} or by ponderomotive scattering in vacuum \cite{koz2021}.

Concerning the chirality of the specimen, we usually have intrinsic chirality in mind, meaning that the studied system lacks mirror symmetry. Plasmonic excitation modes of the system couple free electron wave functions of the same symmetry \cite{zan2019}. This property can be studied in electron energy loss spectroscopy (EELS) schemes \cite{guz2017, ase2014, lou2021} and also applies to dielectric structures as well \cite{kon2023}. Similar to optical CD techniques, the energy loss spectra of electrons may exhibit dichroism with respect to the EVBs OAM. Dichroic signal from free electron beams can also be observed after interacting with near-fields of chiral structures illuminated by opposite circular polarization of light \cite{Har2020}. The chirality can also be imprinted onto the studied system extrinsically, if we consider the circularly polarized light scattering off a sphere. If the spin momentum of the polarized light is transferred to orbital momentum via spin-orbit coupling \cite{bli2015}. This indeed allows for the highly symmetrical sphere to distinguish between left- and right-handed polarization of light due to geometrical effects, giving rise to a chiral near-field.

In this study e propose a scheme, where EVBs spatially overlap with the chiral near-field of nanoparticles. The EVB is sensitive to the symmetry of the near-field, which is shown directly in the discrete EVB energy spectrum measured after quantum coherent interaction with such a chiral near-field.

\section{Theoretical Description}

\begin{figure*}
    \centering
    \includegraphics[width=\textwidth]{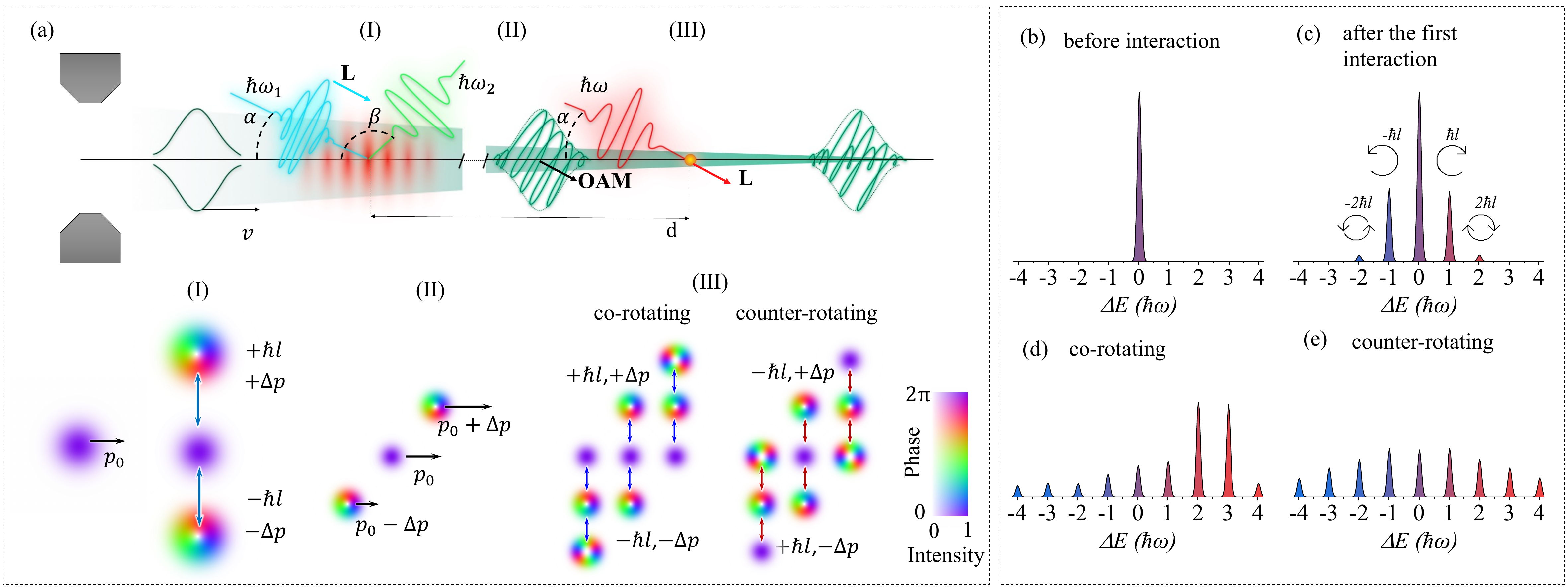}
    \caption{(a) Scheme of the proposed method. (I) A quasi-monochromatic electron beam with Gaussian transverse profile is focused by the lens of the electron microscope. The electron beam scatters off a travelling optical wave created by the interference of a vortex optical wave with photon energy $\hbar\omega_{1}$ and an optical plane wave with photon energy $\hbar\omega_{2}$. In the scattering, the electron beam absorbs photons from one optical field and emits photons into the other, resulting in the generation of side states shifted in energy by $\Delta E= \hbar\omega_{1}-\hbar\omega_{2}$ and by $\Delta p = \hbar (\omega_{1}-\omega_{2})/v$ where each momentum state also carries multiple integer of the OAM of the vortex light field. The transverse dimension of the electron beam is matched to the dimensions of the two laser beams generating the ponderomotive potential. (II) The generated momentum states propagate with different velocities, forming a vortex structure in distance $d$ in the electron wave packet density distribution thanks to the acquired OAM and the electron energy dispersion. The electron beam is focused down to several tens of nanometers before reaching the site of the second interaction. (III) Circularly polarized light field at energy $\hbar\omega$ illuminates a golden nanosphere. The frequency $\omega$ is set to be equal to the difference frequency $\omega_{1}-\omega_{2}$. The spin angular momentum of the illuminating light is converted to OAM of the scattered optical field. The EVB interacts with the scattered chiral near-field. If the OAM of the optical near-field is the same as the OAM of the ponderomotive grating, we call the interaction "co-rotating", and if it is opposite we call it "counter-rotating". The free electron spectra of the (b) initial monochromatic beam prior to modulation, (c) after the first chiral interaction, (d) after two co-rotating chiral interactions, (e) after two counter-rotating chiral interactions.}
    \label{fig:2intscheme}
\end{figure*}

In this section we build up the theoretical basis for the chirality near-field probing. The proposed experimental scheme is shown in Figure \ref{fig:2intscheme}(a). Let us break the method down into three essential steps.

(I) We start with a quasi-monochromatic electron beam with a narrow spectrum (Fig. \ref{fig:2intscheme}(b)). From the initial electron beam focused by the electron microscope optics, we generate a high quality electron vortex beam in a vacuum interaction \cite{koz2021}, which is a superposition of discrete vortex states with different momenta and OAMs. To shape the initial beam into a vortex beam we introduce the first interaction (Fig. \ref{fig:2intscheme}(a)(I)) with two laser beams with photon energies $\hbar\omega_{1}$ and $\hbar\omega_{2}$ propagating under angles $\alpha$ and $\beta$. The $\hbar\omega_{1}$ wave is a vortex wave carrying an OAM $\mathbf{L}$ of magnitude $\hbar l$. The interference pattern of the two beams creates a beat wave which propagates along with the electron wave packet at velocity $v$. While interacting with the beat wave, the electrons absorb and emit photons from the two light fields, leading to energy transfer of integer multiples of $\hbar(\omega_{1}-\omega_{2})$. When we assume the non-recoil approximation (weak interaction regime), the electron spectrum after the first interaction is symmetric, such as in Fig. \ref{fig:2intscheme}(c). Each energy/momentum peak in the free electron spectrum corresponds to a vortex state, which is not only shifted in energy and momentum but also carries an integer multiple of OAM $\hbar l$ absorbed from the optical vortex beam.

(II) At second, we let the superposition of states propagate to a sufficient distance $d$. The probability density of the generated superposition of electron vortex states shifted in momenta reshapes in time and space due to dispersive propagation of electrons in vacuum \cite{dig2020}. After the propagation distance, which is called temporal focus, the probability density in the time domain forms short spikes separated by the period of the modulation fields $T=2\pi/\omega$, where $\omega=\omega_{1}-\omega_{2}$. Thanks to the transverse phase dependence in each vortex state stemming from the OAM transfer, the electron beam probability density is shaped into a spiral in the temporal focus (see Fig. \ref{fig:2intscheme}(a)(II)). In this position we let the electron interact with chiral optical near-fields. The electron vortex dimensions allow to optimize the spatial overlap with the optical near-field at the site of the second interaction.

(III) At third, we introduce a chiral system, in this case a chiral near-field generated around a golden nanosphere (Fig. \ref{fig:2intscheme}(a)(III)), and let the superposition of states interact with it \cite{par2010}. The nanosphere is excited by circularly polarized light with photon energy $\hbar\omega$ with defined and controllable phase difference with respect to the fields in the first interaction.  The sphere acts as a converter between the spin angular momentum of the incident plane wave and the OAM of the scattered chiral field. For left-hand (right-hand) circularly polarized excitation, the scattered field has OAM $+\hbar$ ($-\hbar$). While passing the optical near-field of the sphere, the transitions are induced between the states of the coherent superposition. 

Depending on the OAM of the vortex in the first interaction and the OAM of the near-field of the sphere, two cases are to be expected. If the OAMs have opposite directions (counter-rotating interactions), the spectrum of the electron beam after the two interactions is symmetric (Fig. \ref{fig:2intscheme}(e)) and independent of the relative phase between the modulating optical field and the near-field of the sphere. In contrast, if the OAMs have the same direction (co-rotating interactions), the spectrum of the final electron beam is asymmetric (Fig. \ref{fig:2intscheme}(d)) and the asymmetry depends on the relative phase between the two optical interactions.

%\subsection{\label{sec2:level2:1}Two interaction scheme}

\begin{comment}

\begin{figure}[h]
    \centering
    \includegraphics{FIG2.pdf}
    \caption{Free electron spectra of the a) initial monochromatic beam prior to modulation, b) after single chiral interaction, c) after two counter-rotating chiral interactions, d) after two co-rotating chiral interactions. The final electron wave function consists of coherent superposition of states shifted in energy by integer multiples of $\hbar\omega$ (corresponding momentum shift between neighbouring states is $\hbar\omega/v$ and in OAM by $\hbar l$).}
    \label{fig:spectra}
\end{figure}
\end{comment}

%%% Odstraneno

%The final shape of the spectra is determined by quantum mechanical interference between individual transition amplitudes which depend on the relative phases of discrete electron momentum states acquired during the propagation between the two interactions and on the relative phases of the modulating field and the optical near-field. 
%% Konec odstraneni

\begin{figure}[h!]
    \centering
    \includegraphics{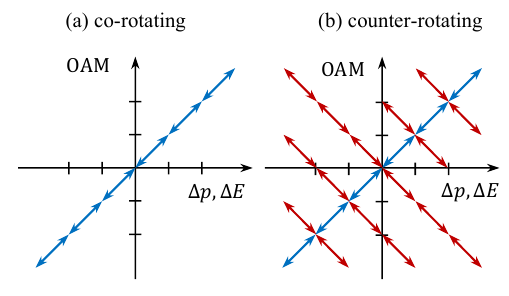}
    \caption{Sketch of the quantum mechanical state interference for co-rotating and counter-rotating interactions. The first interaction creates a ladder of states, where absorption (emission) of momentum also leads to emission (absorption) of OAM. If the interactions have the same OAM, the second interaction leads to redistribution of the state populations along the same ladder of states. If the second interaction has an opposite OAM than the first interaction, in the second interaction the absorption (emission) of momentum leads to emission (absorption) of OAM, which leads to redistribution of the state populations along new state ladders.}
    \label{fig:explanation}
\end{figure}

%%% pridano
From quantum mechanical point of view, the dichroic response comes from the interference of the transition amplitudes of the different momentum states, which are generated from the initially narrow momentum distribution via the phase modulation in the two interaction sites. When the modulation is performed using a vortex ponderomotive potential with OAM $+\hbar$, the OAMs of individual momentum states are separated by $+\hbar$. In the second interaction with the chiral optical near-field, another ladder of momentum states is generated, for which the difference of OAM of the two neighbouring states is $+\hbar$ when the handedness of the circularly polarized excitation is matched to the chirality of the ponderomotive beat wave. In such case, the transition amplitudes interfere because the OAM of each momentum state generated by the second interaction is the same. However, in the case of counter-rotating fields, there is no interference because momentum states with different values of OAM are generated after the second interaction (see the sketch in Fig. \ref{fig:explanation}).

The dichroic response can also be understood from the classical perspective. When the classical ensemble of electrons approximated as point charges interacts with the ponderomotive potential of a beat wave generated by two light waves with planar phase fronts, the momentum component parallel to the electron propagation direction (also its kinetic energy) is modulated \cite{koz2018}. When adding a second spatially shifted phase-locked interaction, the resulting electron energy spectrum changes with the relative phase between the two interactions because the modulated electron density reshapes when propagating between the two interaction sites and forms a periodic train of attosecond pulses \cite{kozak2018}. If the pondoromotive potential in the two interactions would have the same OAM (phase of the interaction parameter is corotating around the electron beam axis in the same direction), the relative phase difference would be constant and the spectra would be modulated as a function of the relative phase after integrating over the transverse electron beam profile. However, in the case of opposite OAMs (counter-rotating phase of the interaction parameter), the integration over the transverse beam profile would smear out any dependence on the relative phase of the two interactions.

%%% konec pridani

%%% pridany obrazek

%%% konec pridani

%% Here I would add the explanation through the quantum path intereference

\subsection{Electron Wave Function Shaping}

Here the proposed experimental scheme is explained in the semiclassical framework, where the free electrons occupy discrete ladder states and the light fields are considered as classical electromagnetic fields. We describe the electron wave function modulation in coordinate representation from which the momentum spectra can be obtained straightforwardly through Fourier transform.

The Hamiltonian describing a non-relativistic electron interacting with electromagnetic field is given as \cite{par2010}
\begin{equation}
\label{eq:01}
    \hat{H} = \frac{1}{2m_{e}}\left( \hat{\mathbf{p}}_{e}-q_{e}\mathbf{A}\right)^{2},
\end{equation}
where $m_{e}$ is the electron mass, $q_e = -e $ is the electron charge, $\hat{\mathbf{p}}_{e}$ is the electron momentum operator and $\mathbf{A}$ is the vector potential. Upon expansion, considering the Coulomb gauge $\nabla\cdot\mathbf{A}=0$, we obtain
\begin{equation}
\label{eq:02}
    \hat{H} = \frac{\hat{\mathbf{p}}_{e}^{2}}{2m_{e}}-\frac{q_e}{m_{e}}\hat{\mathbf{p}}_{e}\cdot\mathbf{A}+\frac{q_{e}^{2}\mathbf{A}^{2}}{2m_{e}}.
\end{equation}
The first term describes free electron propagation in vacuum. The second term describes the coupling between the electron momentum and vector potential and is usually treated as the leading perturbation term for describing electrons interacting with localized near-fields \cite{par2010}. The last term is the so-called ponderomotive potential and becomes significant in vacuum, where the effect of the second term averages up to zero. The gradient of this potential generates a force which repels the electron from high-intensity regions of the light field if the electrons have non-relativistic velocity \cite{PhysRevA.83.063810,axe2020}.

Let us look into the solution of a system with such Hamiltonian. The first term can be dealt with by introducing a suitable interaction picture. The remaining part is then treated as a perturbation denoted $\hat{H}^{\text{ini}}$. The perturbative approach is justified because the change of the electron momentum caused by the interaction is much smaller than it's initial momentum. Based on the discussion provided before, it is enough to consider the dominant contribution, which is given either by the second or by the third term if we consider a field localised around a nanostructure \cite{par2010} or purely vacuum interaction, respectively \cite{koz2021}.

The initial electron wave function is a plane wave $\phi_{\text{ini}}(\mathbf{r},t)$ with energy $E_0=\mathbf{p}_{0}^{2}/2m_{e}$, with a Gaussian temporal envelope $g(\mathbf{r},t)$, giving it a pulsed shape in time and also Gaussian transverse profile $u(\textbf{r})$. Explicitly, we write %pridani u(r)
\begin{equation}
\label{eq:03}
    \phi_{\text{ini}}(\mathbf{r},t)=u(\textbf{r})g(\mathbf{r},t)\exp{\left[ -i\frac{E_0}{\hbar}t+i\frac{\mathbf{p_0}\cdot\mathbf{r}}{\hbar} \right]}.
\end{equation}

According to the path integral representation \cite{sch1981}, the effect of the interaction can be described as an acquired phase factor $e^{i\Phi(\mathbf{r},t)}$, with $\Phi$ defined as
\begin{equation}
\label{eq:04}
    \Phi(\mathbf{r},t)=-\frac{1}{\hbar}\int_{t_{0}}^{t} H^{\text{int}}\left(\mathbf{r}\left(t'\right),t'\right)\,\mathrm{d} t',
\end{equation}
where $\mathbf{r}(t')$ is the classical unperturbed electron trajectory. The wave function of the electron after the interaction is then $\psi(\mathbf{r},t)=e^{i\Phi(\mathbf{r},t)}\phi_{\text{ini}}(\mathbf{r},t)$. From this point on, the discussion slightly differs depending on the leading perturbation term $H^{\text{int}}$ used for the wave function modulation, as we provide a more detailed commentary for each case in the following two sections. However, there is a shared feature irrespective of the kind of interaction, if we consider that either $\mathbf{A}\propto e^{-i\omega t}$ in case of second leading term or $\mathbf{A}^2\propto e^{-i\omega t}$ in case of third leading term. For both cases then the integral yields a product of a constant term $g_{\omega}$ and a sine term that oscillates at frequency $\omega$ (See \cite{par2010,koz2021} respectively for each case). Expansion into plane waves is possible thanks to the sinusoidal phase modulation and mathematically follows from the Anger-Jacobi relation. The modulated wave function written as a superposition of states is %pridani u(r)
\begin{equation}
\label{eq:05}
\begin{split}
&\psi(\mathbf{r},t)=u(\textbf{r})g(\mathbf{r},t)\times\\&\sum_{k=-\infty}^{\infty}\, J_k\left(2|g_\omega|\right)e^{ik\arg{g_\omega}}e^{\left[ -i\left(\frac{E_0}{\hbar}+k\omega\right)t+i\frac{\mathbf{p_{k}}\cdot\mathbf{r}}{\hbar} \right]},
\end{split}
\end{equation}
where $k$ is an integer labeling each energy state and $\mathbf{p}_{k}$ is the momentum corresponding to the energy of the $k$-th side band $E_{0}+k\hbar\omega$. We see that each state is weighted by an amplitude factor given by the $k$-th Bessel function with the argument $2|g_\omega|$, which physically describes the interaction strength. 

Further on we will consider simple propagation of the electron wave packet in the $z$ direction with group velocity $v=|\mathbf{p}_{0}|/m_{e}$. We denote $p_k$ to be the magnitude of the vector $\mathbf{p}_k$. The interaction strength parameter $g_{\omega}(R,\varphi)$ is a function of the transverse coordinates, where $R$ is the distance from the axis and $\varphi$ is the azimuthal angle. Thus we allow for spatial distribution of the probability density of each energy state included in the electron beam. Then the wave function at distance $z$ after the interaction simplifies to
\begin{equation}
\label{eq:06}
\begin{split} %pridano u(R)
    &\psi(R,\varphi,z,t)=u(R)g(t-z/v)\times \\& \sum_{k=-\infty}^{\infty}\, J_k\left(2|g_\omega|\right)e^{ik\arg{g_\omega}}e^{\left[ -i\left(\frac{E_0}{\hbar}+k\omega\right)t+i\frac{p_{k}z}{\hbar} \right]}.
\end{split}
\end{equation}

We further linearly expand the momentum up to the second power of $k$ to account not only for the emergence of momentum states but also for their dispersive propagation \cite{dig2020} 
\begin{equation}
\label{eq:07}
    p_k - p_0 = \hbar k \omega/v - \hbar 2 \pi k^2 /z_{T}+... ,
\end{equation}
where $z_T=4\pi m_{e} v^{3} \gamma^{3} / \hbar \omega^{2}$ is the Talbot distance, with $\gamma$ being the relativistic Lorentz factor \cite{dig2020}. Using the expansion we can factor out the initial wave function and write:
\begin{equation}
\label{eq:08}
\begin{split}
    &\psi(R,\varphi,z,t) =\phi_{\text{ini}}(R,z,t)\times \\ &\sum_{k=-\infty}^{\infty}\, J_k\left(2|g_\omega|\right)e^{ik\arg{g_\omega}}e^{\left[ -ik\omega\left( t -\frac{z}{v}\right) - i2\pi k^2\frac{z}{z_{T}}\right]} .
\end{split}
\end{equation}

To consider a second interaction after propagation distance $d$, we simply treat the wave function obtained in equation (\ref{eq:08}) as the initial wave function at $z=d$. This gives us the expression of the wave function after the second interaction:
\begin{equation}
\label{eq:09}
\begin{split}
   & \psi(R,\varphi,z,t)=\phi_{\text{ini}}(R,z,t) \times\\ &\sum_{k'=-\infty}^{\infty}\, J_{k'} \left[2|g_{\omega'}\left(R',\varphi'\right)|\right]e^{ik\arg{g_\omega'}}e^{\left[ -ik'\omega'\left( t -\frac{z-d}{v}\right) \right]}\times \\ &\sum_{k=-\infty}^{\infty}\, J_k\left[2|g_{\omega}\left(R,\varphi\right)|\right]e^{ik\arg{g_\omega}-ik\omega\left( t -\frac{d}{v}\right) - i2\pi k^2\frac{d}{z_{T}}},
\end{split}
\end{equation}
where $\omega$ is also the modulation frequency of the second interaction, $g'_{\omega}\left(R',\varphi'\right)$ is the interaction strength parameter in the second interaction plane with transverse coordinates $R',\varphi'$, the integer index $k'$ is denoting the energy states generated in the second interaction. The term which is quadratic in $k'$ is omitted, as we are concerned only with the phase relations between the two interactions, and further dispersive propagation after the second interaction will have no effect on the resulting electron spectra. Symbolically we can write
\begin{equation}
\label{eq:10}
    \psi(R,\varphi,z,t)=\mathcal{I}'(\omega)\mathcal{P}(z)\mathcal{I}(\omega)\phi_{ini}(R,z,t),
\end{equation}
where $\mathcal{I}(\omega)$, $\mathcal{I}'(\omega)$ describe the phase modulation after the interactions at the frequency $\omega$, and $\mathcal{P}(z)$ describes the propagation in free space between the sites of the interactions. This propagation distance is usually much larger than the characteristic size of the interaction sites, thus allowing for treating interactions and propagation separately. The electron wave packet spectrum can be obtained from the wave function in momentum representation, which is simply calculated as the Fourier transform in the coordinate representation.

\subsection{Electron Vortex Beam Generation}

%% pictures would be nice
To gain more  specific intuition behind the symbolism introduced in the previous section, we now discuss the case of EVB generation via ponderomotive scattering. As we briefly introduced, the experimental scheme for this interaction comprises of two noncollinear optical beams intersecting the electron beam under angles $\alpha$ and $\beta$ with respect to the $z$ axis defined by the electron beam propagation direction. The optical fields create a beat wave that co-propagates with the electron wave packet at velocity $v$ \cite{koz2018, tsa2023}. One of these beams, however, is an optical vortex with spiralling wavefront \cite{koz2021} with topological charge $l=1$.

To ensure the velocity matching, the angles of the beams $\alpha$, $\beta$, and their angular frequencies $\omega_{1}$, $\omega_{2}$ are tied by the synchronicity condition: \cite{koz2018}
\begin{equation}
\label{eq:11}
    v = \frac{(\omega_{1}-\omega_{2})c}{\omega_{1}\cos{\alpha}-\omega_{2}\cos{\beta}}.
\end{equation}
We further request a combination of angles and frequencies such that the transverse momentum transfer is zero
\begin{equation}
\label{eq:12}
    \omega_{1}\sin{\alpha}-\omega_{2}\sin{\beta}=0.
\end{equation}
This way we ensure the transfer of momentum only in the longitudinal direction, therefore different energy states retain the same direction of propagation as the original beam. 

For simplicity, one can consider the case in which the two optical beams are propagating against each other as described in detail in ref \cite{koz2021}. Here we derive the interaction strength parameter for the more general noncollinear case geometry.

The relevant term of the Hamiltonian is $\frac{e^{2}\mathbf{A}^{2}}{2m_{e}}$. The vector potential can be expressed in terms of the electric intensity $\mathbf{E}=-\partial_{t}A$. The total electric field is given as a superposition of two harmonic fields of the light beams. The complex representation of the fields $\mathbf{E}_{1}$ and $\mathbf{E}_{2}$ both polarized in the $x$ direction are defined as in \cite{koz2021}:
\begin{equation}
\label{eq:13}
\begin{split}
    &\textbf{E}_{1}(\rho_\alpha,\varphi_\alpha,z,t)=\\
    &\frac{\textbf{E}_{10}}{2}g\left(t-\frac{z_{\alpha}}{c}\right)\rho_{\alpha}^{|l|}\times\\
    &\exp{\Big\{-\rho_{\alpha}^{2}-i\Big[\omega_{1}t-\frac{\omega_{1}}{c} z_{\alpha} + il\varphi_{\alpha}\Big]\Big\}},
\end{split}
\end{equation}
\begin{equation}
\label{eq:14}
\begin{split}
    &\textbf{E}_{2}(\rho_\beta,\varphi_\beta,z,t)=\\
    &\frac{\textbf{E}_{20}}{2}g\left(t-\frac{z_{\beta}}{c}\right)\rho_{\beta}^{|l|}\times\\
    &\exp{\Big\{-\rho_{\beta}^{2}-i\Big[\omega_{2}t-\frac{\omega_{2}}{c} z_{\beta}\Big]\Big\}}.
\end{split}
\end{equation}
The real electric fields are described by the real parts. $\textbf{E}_{10}$, $\textbf{E}_{20}$ are the amplitudes of the fields, $\rho_{\alpha}=\sqrt{x_{\alpha}^{2}+y_{\alpha}^{2}}/w_{0}$, $\rho_{\beta}=\sqrt{x_{\beta}^{2}+y_{\beta}^{2}}/w_{0}$ are the radial distances from the beam axes scaled by the beam waist radius $w_{0}$, $\varphi_{\alpha}=\arctan{y_{\alpha}/x_{\alpha}}$ is the azimuthal angle of the optical vortex. Finally the transformed coordinates in the reference frame of the optical beam incident under angle $\alpha$ are $x_{\alpha}=x$, $y_{\alpha}=y\cos{\alpha}-z\sin{\alpha}$, $z_{\alpha}=z\cos{\alpha}+y\sin{\alpha}$. Analogous relations hold for the optical beam incident under angle $\beta$.

Averaging the Hamiltonian over an optical period yields
\begin{equation}
\label{eq:15}
    H^{\text{int}}=\frac{e^2}{4m}\left|\frac{\textbf{E}_{1}}{\omega_{1}}+\frac{\textbf{E}_{2}}{\omega_{2}}\right|^{2}.
\end{equation}
Plugging into Eq. (\ref{eq:04}) we see that the $|\mathbf{E}_{1}|^{2}$ and the $|\mathbf{E}_{2}|^{2}$ terms act on the transverse profile of the beam and cause slight focusing of the electron beam. We consider their effect as multiplication phase factors $e^{i\Phi_{1}(R,\varphi)}$ and $e^{i\Phi_{2}(R,\varphi)}$.
The mixed terms $\mathbf{E}_{1}\cdot\mathbf{E}_{2}^{*}$ and $\mathbf{E}_{1}^{*}\cdot\mathbf{E}_{2}$ have a part oscillating at the difference frequency $\omega$ and therefore the corresponding phase factor is expanded via the Anger-Jacobi relation.

For the acquired phase, considering the phase-matching (\ref{eq:11}) and zero transverse momentum transfer (\ref{eq:12}), the wave function shape right after the first interaction site can be written as:
\begin{equation} %prehozeni clenu ve vzorci
\label{eq:16}
\begin{split}
    &\psi_{1}(R,\varphi,z,t)=\phi_{\text{ini}}(R,z,t)e^{i\Phi_{1}+i\Phi_{2}}\times \\
    &\sum_{k=-\infty}^{\infty}\, \left[J_{k}(2|g^{(1)}_{\omega}|)e^{ik\arg{g^{(1)}_{\omega}}}  e^{ik\frac{\omega}{v}z}e^{-ik\omega t} \right].
\end{split}
\end{equation}
The interaction strength parameter $g^{(1)}_{\omega}$ is: 
\begin{equation}
\label{eq:17}
\begin{split}
    %% version in propagation-distance shifted time
    &g^{(1)}_{\omega}(z)=i\frac{e^{2}E_{pm}^{2}}{4\hbar m \omega_{1}\omega_{2}}\int_{-\infty}^{\infty}\, \mathrm{d}t' \rho_{\alpha}^{|l|}e^{-\rho_{\alpha}^{2}-\rho_{\beta}^{2}}e^{il\varphi_{\alpha}} \times \\
    &g\left(t'-\frac{z_{\beta}}{c}\right)g\left(t'-\frac{z_{\alpha}}{c}\right).
\end{split}
\end{equation}
We can see that $g^{(1)}_{\omega}$ depends on the spatial profile of the optical pulses as well as on their spatio-temporal overlap. The new electron wave function is a superposition of vortex states shifted by $k$ quanta of energy $\hbar(\omega_{1}-\omega_{2})$, momentum $\hbar(\omega_{1}\cos{\alpha}-\omega_{2}\cos{\beta})/c$ and orbital angular momentum $\hbar l$. In collinear beam alignment the $g^{(1)}_{\omega}$ amplitude has a toroid shape in the transverse plane with phase factor $e^{il\varphi}$. Introducing the angles $\alpha$, $\beta$ different from $\alpha=0$ and $\beta=\pi$ squeezes the amplitude profile and tilts the phase. For examples of the calculated $g^{(1)}_{\omega}$ see Supplemental Materials (SM) \cite{SM} Fig. S1.

%added if alpha\neq 0
The momentum modulation is purely longitudinal, so the vortex states do not separate spatially during propagation. However, the transferred OAM retains the direction from the generating optical field, which leads to wavefront tilt if $\alpha\neq 0$. To account for the dispersive propagation and consecutive temporal bunching we include the term quadratic in $k$ as in equation (\ref{eq:07}), giving the wave function propagated to distance $z$ after the interaction. 
\begin{equation}
\label{eq:18}
\begin{split}
    &\psi_{1}(R,\varphi,z,t)=\phi_{\text{ini}}(z,t)e^{i\Phi_{1}+i\Phi_{2}}\times \\
    &\sum_{k=-\infty}^{\infty}\, \big[J_{k}(2|g^{(1)}_{\omega}|)e^{ik\arg{g^{(1)}_{\omega}}}\times \\&  e^{ik\frac{\omega}{v}z}e^{-ik\omega t}e^{ - i2\pi k^2\frac{z}{z_{T}}}
      \big].
\end{split}
\end{equation}

The transverse spatial modulation of the interaction strength parameter $g^{(1)}_{\omega}$ occurs on the scales of the spot sizes of the focused optical beams - several microns, thus the electron beam needs to have similar dimensions in the first interaction site. To probe the near-field in the second interaction site the electron beam needs to be focused down to the transverse size of tens of nanometers. To model this, we consider that the electron beam is convergent already at the first interaction plane, which we describe by introducing a phase factor $\exp{\Big[\frac{ip_{0}(x^{2}+y^{2})}{2\hbar}(\frac{1}{d}-\frac{1}{f})\Big]}$ which leads to the focusing of the beam, with $f$ being the focal distance. The same phase factor is considered for each $k$ state, neglecting the small shifts in momentum. To be able to optimize the spot size in the second interaction plane, we consider the distance $d$ to be different from $f$. The electron wave function in the second interaction plane is obtained via scalar diffraction theory by 2D Fourier transform (denoted by $\mathcal{F}_{\text{2D}}$) of the electron wave function after the first interaction multiplied by the propagation phase. Explicitly, the transformed wave function incident on the second interaction site is
%% added explicit equation
\begin{equation}
\label{eq:S3}
\begin{split}
    &\psi'_{1}(R,\varphi,z=d,t)=\\&\sum_{k=-\infty}^{\infty}\, \mathcal{F}_{\text{2D}}\Bigg\{\phi_{\text{ini}}(R,\varphi,z,t)e^{i\Phi_{1}+i\Phi_{2}} \\& \big[ J_{k}(2|g^{(1)}_{\omega}|)e^{ik\arg{g^{(1)}_{\omega}}}
      \big]\exp{\Bigg[\frac{ip_{0}R^{2}}{2\hbar}\Big(\frac{1}{d}-\frac{1}{f}\Big)\Bigg]}\Bigg\}\\&e^{ik\frac{\omega}{v}d-ik\omega t - i2\pi k^2\frac{d}{z_{T}}}.
\end{split}
\end{equation}

\subsection{Electron Vortex Beams Interacting with Near-Fields}

In the vicinity of a nanostructure illuminated by light, optical near-fields are generated. If the nanostructure is made out of a suitable metal, it acts as a nanoantenna that amplifies the incident lightwave significantly once the light frequency approaches the plasmonic resonances. The structure of the near-field comprises of different electromagnetic modes.

Lets assume the electromagnetic field exciting the nanoparticle to be a plane wave oscillating at the frequency $\omega$, which is incident at angle $\alpha$ with respect to the $z$ axis. The complex representation of the electric field is: 
\begin{equation}
\label{eq:19}
\begin{split}
    &\mathbf{E}_{3}(y,z,t)=\\
    &\frac{\mathbf{E}_{30}}{2}\exp{\Big\{-i\Big[\omega t-\frac{\omega}{c} z_{\alpha}\Big]\Big\}}.
\end{split}
\end{equation}
 The complex amplitude of the field is $\mathbf{E}_{30}=|E_{30}|(\mathbf{\hat{x}}_{\alpha}\pm i\mathbf{\hat{y}}_{\alpha})/\sqrt{2}$, where $\mathbf{\hat{x}}_{\alpha}$ and $\mathbf{\hat{y}}_{\alpha}$ are the unit vectors in the directions $x_{\alpha}$ and $y_{\alpha}$. The next crucial step is to calculate the scattered field, which for the case of a nanosphere can be obtained fully analytically and is descibed by Mie theory \cite{scattering1998}. For more complicated nanostructures, the field has to be modelled numerically.

Following the results in Ref. \citenum{par2010}, for the second interaction placed in $z=d$ we arrive at the wave function modulation given by:
\begin{equation}
\label{eq:20}
\begin{split}
%oprava rovnice
    \psi_{2}(R,\varphi,z,t)=\psi'_{1}(R,\varphi,z=d,t)\times\\
    \sum_{k=-\infty}^{\infty}\, J_k(2|g^{(2)}_{\omega}|)e^{ik\arg{g^{(2)}_{\omega}}}e^{-ik\omega(t-(z-d)/v)}
\end{split}
\end{equation}
where the interaction parameter $g_{\omega}^{(2)}$ is defined as:
\begin{equation}
\label{eq:21}
    g^{(2)}_\omega(R',\varphi') = \frac{e}{\hbar\omega}\int_{-\infty}^{\infty}\,\mathrm{d}z \,E^{(s)}_{z}(R',\varphi',z)e^{i\frac{\omega}{v}z}\,\mathrm{d}z.
\end{equation}
Here $E^{(s)}_{z}$ is the $z$ component of the scattered electromagnetic field. We note that the chirality of the field directly translates into the chirality of $E^{(s)}_{z}$ and thus into the chirality of $g^{(2)}_\omega$. For circularly polarized light with $\alpha=0$ the argument of the coupling parameter $\arg{g_{\omega}^{(2)}}$ is proportional to the angle $\varphi$ with the sign depending on the handedness of the circular polarization of the excitation light wave. For examples of calculated $g^{(2)}_\omega$ for different geometries see SM \cite{SM} Fig. S2.

In the proposed experiment, we set $\omega_{1}=3\omega$ and $\omega_{2}=2\omega$ to match the difference frequency $\omega = \omega_{1}-\omega_{2}$ of the first interaction to the driving frequency of the near-field. The angles of incidence of $\mathbf{E}_{1}$ and $\mathbf{E}_{3}$ are both set to $\alpha$ to match the scattered field chirality to the EVB wavefront tilt. This means that in the second interaction the transitions occur within the discrete momentum states generated in the first interaction.

\subsection{Interpretation of Numerical Simulations}

With these considerations we can step into processing of the numerical simulations. What is to be measured in the proposed experiment are the electron spectra averaged over the entire beam profile. The electron spectrum after the first interaction is given by the Fourier transform of the wave function:
\begin{equation}
\label{eq:22}
\begin{split}
    &|\Psi_{1}(\omega)|^2=\\
    &\left|\int_{R}\mathrm{d}R \int_{\varphi}\mathrm{d}\varphi \int_{-\infty}^{\infty} \mathrm{d}t\,\psi_{1}(R,\varphi,z,t) e^{i2\pi\omega t}\right|^2.
\end{split}
\end{equation}
The initial single peak spreads into a symmetric comb of energy side bands (see Fig. \ref{fig:2intscheme}(c)). Even though the spectra vary within the transverse plane because the amplitudes follow $J_k(2|g_\omega|)$ which is not homogeneous, all spectra are symmetrical with respect to the central energy.

The spectrum after the second interaction is obtained analogously
\begin{equation}
\label{eq:23}
\begin{split}
   & |\Psi_{2}(\omega)|^2=\\
   &\left|\int_{R}\mathrm{d}R \int_{\varphi}\mathrm{d}\varphi \int_{-\infty}^{\infty} \mathrm{d}t\,\psi_{2}(R,\varphi,z,t) e^{i2\pi\omega t}\right|^2.
\end{split}
\end{equation}
However, here we ought to distinguish two separate cases. It shows, that if the EVB and the near-field are perfectly counter rotating, the mutual phase relations within the transverse beam profile compensate each other with opposite signs and their effect evens out, leading again to a symmetrical spectrum as schematically shown in Fig. \ref{fig:2intscheme}(d). In contrast, if the EVB and the near-field have the same chirality, the produced average spectrum is non-symmetrical as the phase differences are maintained within the transverse plane and no longer even each-other out as illustrated in Fig.\ref{fig:2intscheme}(e). This is precisely true for the case in which the two modulating optical beams are collinear, the electron vortex beams have untilted wavefronts and the near-field of the nanoobject has cylindrical symmetry. In the next section we investigate the behavior of such system, however we also test the method robustness for more realistic cases as well. For detailed semi-classical explanation of the symmetry/asymmetry of the free electron spectra see SM \cite{SM}.

To quantify the asymmetry $A$ or symmetry $S$ of the spectra, we define the following parameters:
\begin{equation}
\label{eq:24}
\begin{split}
   & \tilde{A} = \sqrt{\sum_{k}\frac{1}{2}\Big[|\Psi(k\omega)|^2-|\Psi(-k\omega)|^2\Big]^{2}},\\
   & \tilde{S} = \sqrt{\sum_{k}\frac{1}{2}\Big[|\Psi(k\omega)|^2+|\Psi(-k\omega)|^2\Big]^{2}},\\
\end{split}
\end{equation}
with the normalization:
\begin{equation}
\label{eq:25}
\begin{split}
   & A = \frac{\tilde{A}}{\tilde{A}+\tilde{S}},\\
   & S = \frac{\tilde{S}}{\tilde{A}+\tilde{S}}.\\
\end{split}
\end{equation}
The asymmetry and symmetry are then complementary values, because $A+S=1$. The $A$ parameter can be taken as a function of the mutual interaction phase, that is varied and calculated for each spectrum separately.

\section{Simulation Results}

\begin{figure*}
    \centering
    \includegraphics[width=\textwidth]{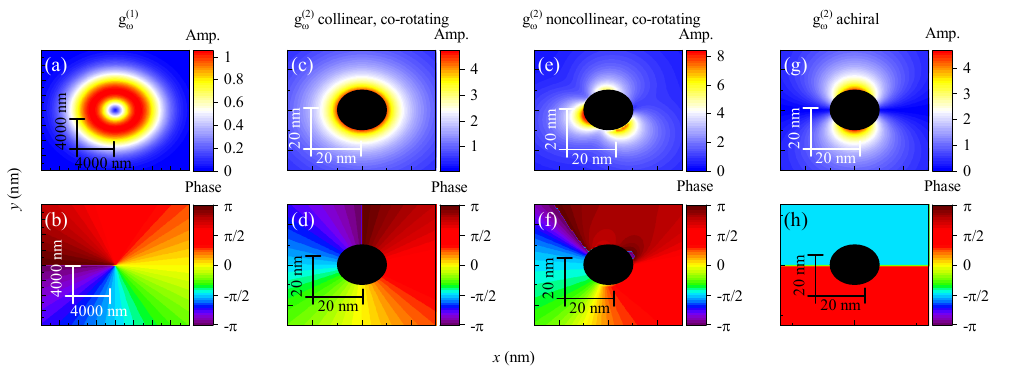}
    \caption{Examples of the amplitude and phase of the interaction parameters $g^{(1,2)}_{\omega}$. The interaction parameter $g_{\omega}^{(1)}$ (a,b) in the first interaction site has a doughnut amplitude profile and a phase proportional to the azimuthal angle $\varphi$ for OAM $l=1$. The parameter $g_{\omega}^{(2)}$ in the second interaction site is shown for (c,d) co-rotating, collinear excitation of the nanosphere, (e,f) co-rotating, non collinear excitation (g,h) example of achiral (linearly polarized) excitation. The black circle in (c-h) signifies the geometric shadow of the nanosphere.}
    \label{fig:gparams}
\end{figure*}

The frequencies of the optical beams used for the first interaction are $\omega_{1}=3.6\,\mathrm{eV}$, $\omega_{2}=2.4\,\mathrm{eV}$. The angles are $\alpha=0^{\circ}$, $\beta=180^{\circ}$ for collinear geometry, for which the free electrons are phase matched to the beam wave when they have velocity of $v=0.20c$.  For noncollinear geometry we use the angles $\alpha=13.18^{\circ}$, $\beta=160.00^{\circ}$ to create a light wave phase-matched with electrons at velocity of $v=0.21c$ with zero transverse momentum transfer. For simplicity, we consider the same  $e^{-2}$ beam waist of radius $w_0=5\,\mathrm{\mu m}$, pulse duration of $\tau_\text{opt}=500\,\mathrm{fs}$ and the amplitude selected such that the maximum of the interaction strength parameter is close to 1 ($\text{max}(|g_{\omega}^{(1)}|)=1.04$ in the collinear case and $\text{max}(|g_{\omega}^{(1)}|)=0.87$ in the noncollinear case), which give strong first photon sidebands and minimize zero loss peak. The calculated shape of the $g^{(1)}_{\omega}$ parameter for colinear geometry and $l=1$ is shown in Fig. \ref{fig:gparams}(a,b).

The electron beam enters the first interaction with a Gaussian transverse profile with the $e^{-2}$ radius of $w_{e}=3.5\,\mathrm{\mu m}$ and FWHM duration of the wave function envelope $\tau_{\text{el}}=200\,\mathrm{fs}$. The second interaction is placed within the temporal focus, at propagation distance of $d=0.658\,\mathrm{mm}$, where the density of the phase-modulated vortex electron beam restructures and forms attosecond bunches separated by the period of the modulating optical field ($T=2\pi/\omega$). To ensure optimal size of the electron beam at the site of the second interaction, we simulate the focusing of the electron beam with a lens with focal distance $f =0.7\,\mathrm{mm}$, corresponding to $5\,\mathrm{m rad}$ beam convergence by adding a spherical phase term. For those parameters, the initial Gaussian beam is focused to a $e^{-2}$ radius of approximately $24\,\mathrm{nm}$ at the second interaction site.

\subsection{Plasmonic Nanosphere}

A golden nanosphere is a suitable prototype to demonstrate the concept of the method because the scattered near-field has OAM and can be easily calculated analytically. The optical near-fields in the vicinity of the nanosphere with the radius of $10\,\mathrm{nm}$ are calculated using Mie theory with the dielectric function of gold described in the framework of the Lorenz-Drude model \cite{Rakic:98}. The frequency of the illuminating field is $\omega=1.2\,\mathrm{eV}$. With $E^{(s)}_{z}$ directly plugged into (\ref{eq:21}) we obtain $g_{\omega}^{(2)}$ with a doughnut-shaped intensity profile and azimuthaly varying phase.

In Fig. \ref{fig:gparams}(c,d) is the calculated $g^{(2)}_{\omega}$ for left-hand circularly polarized illumination of the golden nanosphere, incident along the $z$ axis. We note that the phase (d) has the same sense of rotation as the phase of the ponderomotive potential (b), hence the term “co-rotating, collinear geometry”. For the counter-rotating case the phase would have inverted sense of rotations (See SM \cite{SM} Fig. S2). In (e,f) the nanosphere is illuminated by left-hand circularly polarized plane wave incoming from angle $\alpha=13.18^{\circ}$ with respect to the $z$ axis, to which we refer as “co-rotating, noncollinear” geometry. In (g,h) the nanosphere is illuminated by linearly polarized light along the $z$ axis, resulting in a achiral $g_{\omega}^{(2)}$.

The amplitude of the wave is set to give $max(|g_{\omega}^{(2)}|)\approx4.4$ outside of the sphere for the collinear case or $max(|g_{\omega}^{(2)}|)\approx8$ for the noncollinear case, which results in the generation of up to 6 side bands after two interactions in the counter-rotating case. Plugging into Eq. (\ref{eq:20}), we imprint the phase of $|g_{\omega}^{(2)}|$ onto the wave function.

%%%% tady bych vlozila obrazek g parametru

\subsubsection{Collinear and noncollinear geometry}

\begin{figure}[ht]
    \centering
    \includegraphics{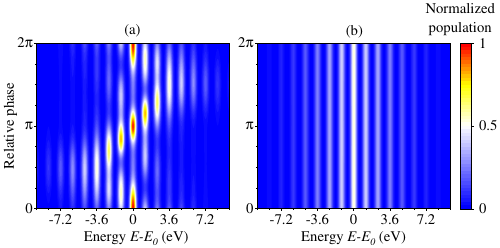}
    \caption{Electron spectra of electron vortex beam (coherent superposition of momentum-shifted vortex states) generated via ponderomotive scattering in collinear geometry after propagation and interaction with optical near-fields of a golden nanosphere illuminated by circularly polarized light with the handedness of polarization matched to the direction of the electron orbital angular momentum (a) and opposite to it (b).}
    \label{fig:03}
\end{figure}

First we focus on the case with collinear geometry. The spectra calculated through Eq. (\ref{eq:23}) are plotted in Figure \ref{fig:03}. We observe that in the co-rotating case the spectra are strongly asymmetric and the populations of individual k-states oscillate as a function of the relative phase between the two interactions. In contrast, for the counter-rotating case we observe phase-independent symmetrical profile of the spectra.

\begin{figure}[ht]
    \centering
    \includegraphics{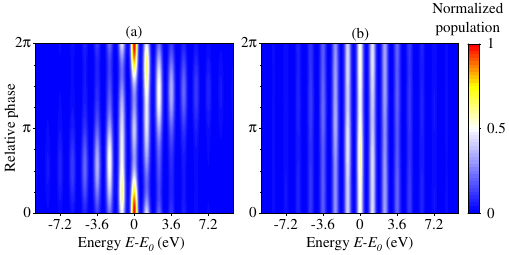}
    \caption{Electron spectra of electron vortex beam (coherent superposition of momentum-shifted vortex states) generated via ponderomotive scattering in noncollinear geometry with angles  $\alpha=13.18^{\circ}$ and $\beta = 160.00^{\circ}$ after propagation and interaction with optical near-fields of a golden nanosphere illuminated by circularly polarized light under angle  $\alpha=13.18^{\circ}$ with the handedness of polarization matched to the direction of the electron orbital angular momentum (a) and opposite to it (b).}
    \label{fig:04}
\end{figure}

The experimental realization of the collinear geometry is complicated due to the necessity of the use of mirrors with holes for the electron beam, which can distort the field profile of light beams generating the electron vortex beam. For this reason we simulate a more realistic case, which allows for experimental realization. In the simulation of this noncollinear scheme we use the angles $\alpha=13.18^{\circ}$ and $\beta=160.00^{\circ}$. In Fig. \ref{fig:04} we plot the resulting spectra, which slightly differ from the collinear geometry case. The asymmetry of the spectra is less pronounced in the co-rotating case, albeit still present. In the counter-rotating case, a slight asymmetry appears with the variation of the mutual phase. This is expected as the interaction strength parameters are no longer symmetric around the axis of electron propagation.

\begin{figure}[ht]
    \centering
    \includegraphics{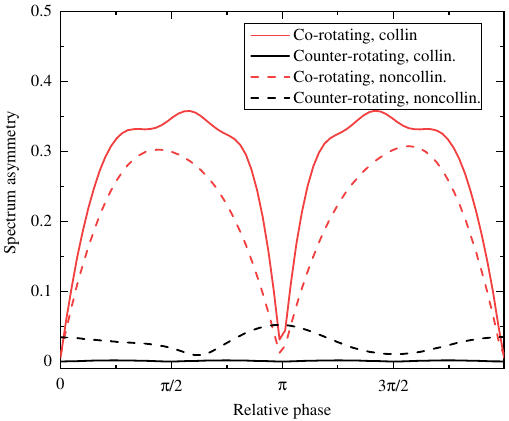}
    \caption{Spectrum asymmetry parameter for collinear (full line) and noncollinear (dashed line) geometry of the chirality probing. For co-rotating interactions (red) we see significant modulation of the asymmetry with the changing mutual phase. For counter-rotating interactions (black) the asymmetry remains low and varies slowly.}
    \label{fig:05}
\end{figure}

The difference can be characterized by the spectrum asymmetry parameter defined by Eq. (26), which is plotted in Fig. \ref{fig:05} for all possible mutual phases of the interactions. The asymmetry for co-rotating configuration (red lines) varies strongly with the phase. 
When the mutual phase between the two interactions is $0$ or $\pi$, the spectra are symmetric even for the co-rotating case (see the dips in the red solid curve shown in Fig. \ref{fig:05}). For counter-rotating configuration (black curves), the asymmetry is lower.

Note that for distinguishing the chirality of the near-field it is necessary to measure the electron spectra for both electron vorticities, $l=\pm 1$. The asymmetric and phase dependent electron spectra for a given vorticity alone are not a sufficient indicator of the chirality of the near-field. Let us assume a simple example of an achiral system, e.g., plasmonic nanosphere illuminated by linearly polarized light. The structure of $g_{\omega}^{(2)}$ constitutes of two lobes with opposite signs (see Fig. \ref{fig:gparams}(g,h)), so when the electron vortex superposition interacts with the near-field, the asymmetrical spectra emerging on the opposite sites in the electron beam around the sphere do not even out because the geometrical phase of the vortex states is compensated by the geometrical phase of the near-field. This leads again to compounding of asymmetric spectra in the averaging process. This asymmetry is, similarly to the chiral case, phase sensitive. However switching the electron beam vorticity yields the same result, because the interaction does not differentiate between the two OAMs. Ultimately, to determine the chirality of a near-field, a dichroic response is required.

\subsubsection{Off-axis alignment}
We further analyze how the electron spectra behave once we introduce a misalignment between the two interactions and the sphere is no longer positioned on the axis of the electron beam. The asymmetry is expected to decrease for the co-rotating configuration and increase for the counter-rotating configuration because of breaking the axial symmetry. The question remains how far from the structure the electron beam can be positioned to still resolve the chirality of the field. For simplicity we assume the collinear geometry. In collinear alignment, we neglect the spectra of the part of the electron beam that would have passed through the sphere because most of these electrons inelastically scatter. 

\begin{figure}[ht]
    \centering
    \includegraphics{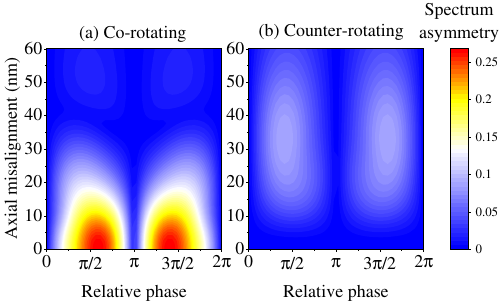}
    \caption{Electron beam spectrum asymmetry as a function of the relative phase between the interactions and the transverse distance between the axis of the electron beam and the position of the plasmonic nanosphere for collinear geometry, (a) co-rotating and (b) counter-rotating case.}
    \label{fig:06}
\end{figure}

The results are shown in Figure \ref{fig:06}. The resulting spectra when scanning the sphere near field are affected by the electron beam and near-field overlap, concerning both the phases and the amplitudes. For the co-rotating case we see gradual decrease in the asymmetry parameter $A$, dropping to half it's maximal value around $20\,\mathrm{nm}$ axial distance as the phase symmetry breaks and the overlap is distorted.

Similarly, for counter-rotating configuration the asymmetry parameter starts as zero for perfectly aligned position and as the phases begin to drift, the spectral profiles do not even out on average, the asymmetry starts to rise. Around $30\,\mathrm{nm}$ a there is a peak in the asymmetry, thanks to the partially restored overlap and phase relations, allowing for some asymmetry to appear.

This effect suggests that the electron vortex superposition is sensitive to the local near-field chirality even when the overlap with the field is just partial. However, it also complicates the asymmetry signal analysis, since the contribution from the near-field and the electron beam are convoluted together. Scanning for the asymmetry integrated over all interaction phases in all directions in the vicinity of the sphere would yield a central region with strong asymmetry for the co-rotating beam surrounded by a ring of lower asymmetry for the counter-rotating beam. The dimensions of the features ultimately depend on the size of the nanostructure and the size and the shape of the electron beam, which can be used to optimize the resolution. 

\subsubsection{Effect of overlap with the optical fields and longitudinal and transverse coherence}
%% tohle cele je pridane

In a realistic situation, the real space dimensions of the electron pulse and the interaction regions and the coherence properties of the electrons play a role. 

In the first interaction we choose the laser pulse duration to be $\tau_{\text{opt}}=500\,\mathrm{fs}$ to ensure homogeneous amplitude of the modulation across the electron time profile. In our geometry and with a low electron velocity of $v = 0.2c$, the optical pulse duration approximately corresponds to the effective interaction time. The effective interaction length in the $z$-direction can be obtained by multiplying interaction time $\tau$ by the speed of light $c$, yielding $150\,\mathrm{\mu m}$, while the spatial length of the electron wave packet corresponds to its duration multiplied by its velocity $v$ and is equal to $12\,\mathrm{\mu m}$.  Because the interaction length is significantly larger than the length of the electron wave packet, the amplitude of the electron energy modulation is practically homogeneous within the electron wave packet envelope along the propagation direction $z$.

On the other hand, the ponderomotive grating oscillation has a period of $T=3.44\,\mathrm{fs}$ (corresponding to the difference energy $\hbar\omega_{1}-\hbar\omega_{2}=1.2\,\mathrm{eV}$). For the spectral peaks to be distinguishable, the electron coherence time needs to be long enough for the coherent part of the wave function to capture the modulation period of the ponderomotive grating. We consider initial width of the electron peak $\Delta E = 0.5\,\mathrm{eV}$ which is within the range of typical energy widths for common pulsed electron sources \cite{Feist2016, Yannai2023}. This corresponds to coherence time of approximately $\tau_{\text{coh}}=3.64\,\mathrm{fs}$ FWHM, which was sufficient to observe the quantum interference between electron transition amplitudes in previous experiments \cite{fei2015}. Further broadening of the energy spectrum and shortening of the coherence time relative to the modulation period smears the spectra and the side peaks cannot be resolved anymore. However the spectral modulation is still present \cite{kozak2018}. This can be understood even in the classical framework, where the electrons experience energy modulation thanks to the Lorentz force. The modulated electron beam re-shapes and forms a narrow spikes in the longitudinal density distribution correspondig to an attosecond pulse train. All pulses from the train come to the second interaction with constant time delay with respect to the field cycles of the optical near-field. The effect of the energy modulation is expected to vanish if the initial electron energy spectrum is wider than the spectral broadening induced by the optical field modulation.

The second inelastic scattering of the electrons occurs via the interaction with the chiral optical near-field of a nanosphere illuminated by coherent circularly polarized light. Here we assume a constant amplitude of the incident light, which is used as an approximation of illumination by a pulse with pulse duration much longer than the duration of the electron wave packet. The frequency of the circularly polarized light is the same as the modulation frequency of the first interaction resulting in the same distance between the generated photon side bands.

Computation-wise, we describe the interaction of a fully coherent electron pulse. The finite temporal coherence results only in broadening of the spectra and it can be modelled in two ways - the narrow spectra can be broadened through convolution, simulating the smearing due to incoherent summing of spectra shifted in energy, or the wave function can be multiplied by a short Gaussian time window corresponding to the coherence time prior to calculating the spectra. Both approaches are mathematically equivalent because the initial width of electron energy spectrum is small and the interaction parameters $g^{(1,2)}_{\omega}$ do not depend on the electron energy within the initial spectrum.

The proposed experiment relies on stable phase relations between the electron states of the electron beam and the probed near-field. The lack of transverse coherence might be detrimental to the effect. However, transverse coherence high enough to generate the vortex states in the far field can be achieved as it was proven in other experiments demonstrating generation of electron vortex beams \cite{ver2010}.

\subsection{Intrinsically Chiral Nanoantennae}

In this section we present several chiral nano-structures to illustrate the possibility of chirality probing on artificial chiral fields generated by scattering linearly polarized light. The idea is that even though the excitation light itself is not circularly polarized or in a vortex form, the spatial distribution of the field scattered from the structure acquires such phase relations, that it behaves as if it was chiral. We show that free electron vortices can also be used to study such special fields.

\begin{figure}[ht]
    \centering
    \includegraphics[]{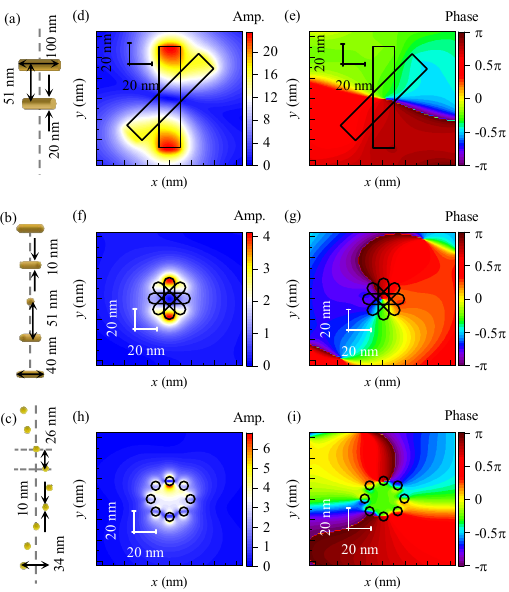}
    \caption{Nanostructures with intrinsic chirality: a) two rods, b) nanorod helix, c) nanosphere helix. Near-field complex interaction parameter $g_{\omega}^{(2)}$ of the (d,e) two rod structure, (f,g) nanorod helix, (h,i) nanosphere helix. The area in the black outline in (d-i) signifies the geometrical shadow of the nanostructures. }
    \label{fig:07}
\end{figure}

We investigate 3 different nanostructure geometries based off golden nanoparticles. In Fig. \ref{fig:07}(a) we show the simplest geometry which consists of two rods with $a = 20\,\mathrm{nm}$ diameter and $b = 100\,\mathrm{nm}$ length separated by $s = 51\,\mathrm{nm}$ in the $z$ direction and rotated by $\pm 45^{\circ}$ with respect to each other. Then we consider two helical geometries inspired by previously manufactured nanoscale antennae \cite{kuzyk2012, lan_au_2015}. In Fig. \ref{fig:07}(b) we show a model of a helix made out of nanorods with $a=10\,\mathrm{nm}$ diameter and $b=40\,\mathrm{nm}$ and $s = 51\mathrm{nm}$ spacing in the $z$ direction and $45^{\circ}$ rotation between the segments. In Fig. \ref{fig:07}(c) we show a model of a helix made out of nanospheres with $a=10\,\mathrm{nm}$ diameter and $r=17\,\mathrm{nm}$ distance from the $z$ axis and $s = 26 \mathrm{nm}$ spacing in the $z$ direction and rotations of $45^{\circ}$. In the simulation we neglect the support of the individual parts, however, for practical applications the material needs to be considered carefully, to avoid unwanted material charging and electron scattering. The contribution from electrons in the shadow of the nanostructures is filtered out.

In these calculations we consider the collinear geometry of the modulating laser beams and linear polarization along the $y$ axis of the light wave exciting the near-fields of the nanostructure with photon energy of  $\hbar\omega=1.2\,\mathrm{eV}$, incident on the nanostructure along the $z$ direction. The figure of merit to help us quantify the sensitivity to the chirality of the scattered field remains $\arg{g_{\omega}^{(2)}}$ analogously to the sphere case. The nanostructures allows to engineer $g_{\omega}^{(2)}$ through variations of their parameters.  The electric fields are calculated  using finite-difference time-domain technique. The amplitude and phase of the calculated $g_{\omega}^{(2)}$ parameter is shown in Fig. \ref{fig:07}(d-i).

\begin{figure}[ht]
    \centering
    \includegraphics[]{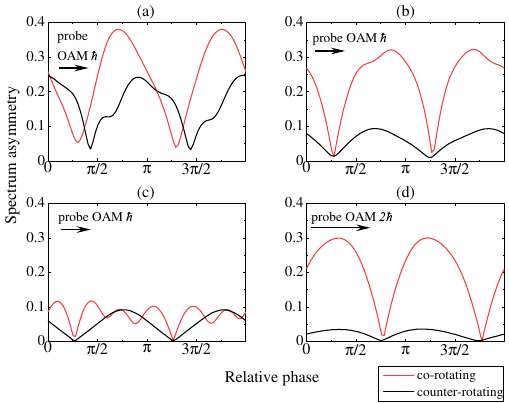}
    \caption{Spectra asymmetry for intrinsically chiral nanostructures: (a) two rods (b) nanorod helix (c) nanosphere helix probed with OAM $\hbar$, (d) nanosphere helix probed with OAM $2\hbar$.}
    \label{fig:08}
\end{figure}

The asymmetries in the electron spectra measured after the scattering of vortex electron beam at the near-fields of the chiral nanostructures are shown in Figure \ref{fig:08}. The resulting spectra and their asymmetry are mainly influenced by the OAM purity of the generated near-field (for the calculated OAM spectra see SM \cite{SM} Fig. S4 (f,g,h)). The simplest two nanorod geometry exhibits the weakest dichroic response. From the $g_{\omega}^{(2)}$ phase in Fig. \ref{fig:07}(e) it is visible that the the near-field has poor OAM purity. In contrast the nanorod helix structure near-field has very high purity of OAM $\pm\hbar$ (Fig. \ref{fig:07}(g)). For the given parameters, the nanosphere helix near-field (Fig. \ref{fig:07}(i)) has strong contribution of OAM $\pm2\hbar$, thus is expected to have stronger response when probed with electron states modulated with OAM $2\hbar$ in the ponderomotive interaction.

\section{Discussion}

The proposed method relies conceptually on
quantum state mixing of the free electron vortex
states. The spectral properties depend on the phase relations between the electron states within the beam. The mixing is influenced by the topology of the near-field. To study only phase effects, a homogeneous amplitude of the interaction strength parameters around the $z$ axis is required. However, introducing angles inherently disrupts the axial symmetry of the interaction parameter amplitudes. Perfect conditions would therefore call for collinearity of all beams within the experiment. Nevertheless, keeping the collinear alignment within the first interaction is more challenging and would require the use of hollow mirrors to let the electrons pass. Additionally, collinear geometry reduces the degrees of freedom that can be adjusted to phase-match the light wave to higher electron velocities and then to couple to different modes within the near-field of the studied nanoparticle.

As we show, even with quite large angles it is still possible to distinguish the chirality of the optical near-field. What comes into question then is the free space propagation of electron states with OAM which is not collinear with their momentum. Alternatively, different approach can be used to generate the electron vortex state superposition using optical or plasmonic near-fields \cite{van2019}, provided that all the vortex states propagate collinearly and that the energy levels can be sufficiently resolved.

Recently, methods for reconstructing both the amplitude and the phase of near-fields were demonstrated. The complete optical near-field can be imagined by Lorentz-PINEM \cite{gaida2023}, free-electron homodyne detection (FREHD) \cite{gaida2023xiv} or in attosecond field-cycle-contrast electron microscopy \cite{nab2023}. Allowing for the full reconstruction of the near-fields, their chirality can be also measured from the local phase calculations. The method proposed in this article, similarly to the latter two, relies on the phase locking of two consecutive coherent electron interactions, and can be considered as yet another alternative approach suitable for chiral fields. 

\section{Conclusions}

In this study we introduce a free electron vortex dichroism method for nanoscale chirality probing using a coherent superposition of vortex electron states, in principle requiring only measurements of the overall beam electron spectra and their dependence on the phase between the modulating optical fields and the near-field of the investigated nanostructure. The electron vortex state superposition can be created in vacuum via ponderomotive scattering to preserve the quality of the electron beam and to prevent elastic and inelastic scattering of electrons at the site where the modulation is introduced. The sensitivity of the vortex superposition to the near-field chirality is demonstrated on two model systems - plasmonic near-field of a golden nanosphere excited by circularly polarized light (extrinsic chirality) and near-fields of chiral golden nanoantenae excited by linearly polarized light (intrinsic chirality). We show that the chirality of the near-field manifests in the symmetry or asymmetry of the electron vortex state superposition spectra after the interaction with the chiral near-fields. The spectrum symmetry can be characterized using dimensionless symmetry and asymmetry parameters which can serve for general discrimination of dichroic behaviour. This behavior is investigated under different circumstances (introducing angles between the OAM and the $z$ axis, scanning the vicinity of the nanoscatterer with the electron beam).

We show that the proposed method is robust with respect to noncollinearity, axial misalignment and can be used to determine the chirality of near-fields of single nanoparticles with nanometre spatial resolution that can be optimized to the particle sizes thanks to the electron beam focusability. When considering the atomic resolution of state-of-the-art transmission electron microscopes, this method may be in future applied to image the chiral optical response of single molecules.

Note added. During the review process we became aware of the work of Fang et al. \cite{fang2024} where the authors demonstrated experimentally the use of electron vortex beams for chirality measurement.

\section*{Acknowledgement}
The authors thank Tomáš Novotný and Tomáš Ostatnický for consultations and fruitful discussions.\\
The authors acknowledge funding from the Czech Science Foundation (project 22-13001K), Charles University (SVV-2023-260720, PRIMUS/19/SCI/05, GAUK 90424) and the European Union (ERC, eWaveShaper, 101039339). Views and opinions expressed are however those of the author(s) only and do not necessarily reflect those of the European Union or the European Research Council Executive Agency. Neither the European Union nor the granting authority can be held responsible for them.
This work was supported by TERAFIT project No. CZ.02.01.01/00/22\_008/0004594 funded by OP JAK, call Excellent Research.

\providecommand{\noopsort}[1]{}\providecommand{\singleletter}[1]{#1}%


\begin{thebibliography}{52}%
\makeatletter
\providecommand \@ifxundefined [1]{%
 \@ifx{#1\undefined}
}%
\providecommand \@ifnum [1]{%
 \ifnum #1\expandafter \@firstoftwo
 \else \expandafter \@secondoftwo
 \fi
}%
\providecommand \@ifx [1]{%
 \ifx #1\expandafter \@firstoftwo
 \else \expandafter \@secondoftwo
 \fi
}%
\providecommand \natexlab [1]{#1}%
\providecommand \enquote  [1]{``#1''}%
\providecommand \bibnamefont  [1]{#1}%
\providecommand \bibfnamefont [1]{#1}%
\providecommand \citenamefont [1]{#1}%
\providecommand \href@noop [0]{\@secondoftwo}%
\providecommand \href [0]{\begingroup \@sanitize@url \@href}%
\providecommand \@href[1]{\@@startlink{#1}\@@href}%
\providecommand \@@href[1]{\endgroup#1\@@endlink}%
\providecommand \@sanitize@url [0]{\catcode `\\12\catcode `\$12\catcode
  `\&12\catcode `\#12\catcode `\^12\catcode `\_12\catcode `\%12\relax}%
\providecommand \@@startlink[1]{}%
\providecommand \@@endlink[0]{}%
\providecommand \url  [0]{\begingroup\@sanitize@url \@url }%
\providecommand \@url [1]{\endgroup\@href {#1}{\urlprefix }}%
\providecommand \urlprefix  [0]{URL }%
\providecommand \Eprint [0]{\href }%
\providecommand \doibase [0]{https://doi.org/}%
\providecommand \selectlanguage [0]{\@gobble}%
\providecommand \bibinfo  [0]{\@secondoftwo}%
\providecommand \bibfield  [0]{\@secondoftwo}%
\providecommand \translation [1]{[#1]}%
\providecommand \BibitemOpen [0]{}%
\providecommand \bibitemStop [0]{}%
\providecommand \bibitemNoStop [0]{.\EOS\space}%
\providecommand \EOS [0]{\spacefactor3000\relax}%
\providecommand \BibitemShut  [1]{\csname bibitem#1\endcsname}%
\let\auto@bib@innerbib\@empty
%</preamble>
\bibitem [{\citenamefont {Johnson}(1990)}]{joh1990}%
  \BibitemOpen
  \bibfield  {author} {\bibinfo {author} {\bibfnamefont {W.~C.}\ \bibnamefont
  {Johnson}},\ }\bibfield  {title} {\bibinfo {title} {Protein secondary
  structure and circular dichroism: A practical guide.},\ }\href@noop {}
  {\bibfield  {journal} {\bibinfo  {journal} {Proteins}\ }\textbf {\bibinfo
  {volume} {7}},\ \bibinfo {pages} {205–214} (\bibinfo {year}
  {1990})}\BibitemShut {NoStop}%
\bibitem [{\citenamefont {B\"owering}\ \emph {et~al.}(2001)\citenamefont
  {B\"owering}, \citenamefont {Lischke}, \citenamefont {Schmidtke},
  \citenamefont {M\"uller}, \citenamefont {Khalil},\ and\ \citenamefont
  {Heinzmann}}]{bow2001}%
  \BibitemOpen
  \bibfield  {author} {\bibinfo {author} {\bibfnamefont {N.}~\bibnamefont
  {B\"owering}}, \bibinfo {author} {\bibfnamefont {T.}~\bibnamefont {Lischke}},
  \bibinfo {author} {\bibfnamefont {B.}~\bibnamefont {Schmidtke}}, \bibinfo
  {author} {\bibfnamefont {N.}~\bibnamefont {M\"uller}}, \bibinfo {author}
  {\bibfnamefont {T.}~\bibnamefont {Khalil}},\ and\ \bibinfo {author}
  {\bibfnamefont {U.}~\bibnamefont {Heinzmann}},\ }\bibfield  {title} {\bibinfo
  {title} {Asymmetry in photoelectron emission from chiral molecules induced by
  circularly polarized light},\ }\href
  {https://doi.org/10.1103/PhysRevLett.86.1187} {\bibfield  {journal} {\bibinfo
   {journal} {Phys. Rev. Lett.}\ }\textbf {\bibinfo {volume} {86}},\ \bibinfo
  {pages} {1187} (\bibinfo {year} {2001})}\BibitemShut {NoStop}%
\bibitem [{\citenamefont {Beaulieu}\ \emph {et~al.}(2018)\citenamefont
  {Beaulieu}, \citenamefont {Comby},\ and\ \citenamefont {Descamps}}]{bea2018}%
  \BibitemOpen
  \bibfield  {author} {\bibinfo {author} {\bibfnamefont {S.}~\bibnamefont
  {Beaulieu}}, \bibinfo {author} {\bibfnamefont {A.}~\bibnamefont {Comby}},\
  and\ \bibinfo {author} {\bibfnamefont {D.~e.~a.}\ \bibnamefont {Descamps}},\
  }\bibfield  {title} {\bibinfo {title} {Photoexcitation circular dichroism in
  chiral molecules},\ }\href@noop {} {\bibfield  {journal} {\bibinfo  {journal}
  {Nat. Phys.}\ }\textbf {\bibinfo {volume} {14}},\ \bibinfo {pages}
  {484–489} (\bibinfo {year} {2018})}\BibitemShut {NoStop}%
\bibitem [{\citenamefont {Ayuso}\ \emph {et~al.}(2021)\citenamefont {Ayuso},
  \citenamefont {Ordonez}, \citenamefont {Ivanov},\ and\ \citenamefont
  {Smirnova}}]{Ayuso2021}%
  \BibitemOpen
  \bibfield  {author} {\bibinfo {author} {\bibfnamefont {D.}~\bibnamefont
  {Ayuso}}, \bibinfo {author} {\bibfnamefont {A.~F.}\ \bibnamefont {Ordonez}},
  \bibinfo {author} {\bibfnamefont {M.}~\bibnamefont {Ivanov}},\ and\ \bibinfo
  {author} {\bibfnamefont {O.}~\bibnamefont {Smirnova}},\ }\bibfield  {title}
  {\bibinfo {title} {Ultrafast optical rotation in chiral molecules with
  ultrashort and tightly focused beams},\ }\href
  {https://doi.org/10.1364/OPTICA.423618} {\bibfield  {journal} {\bibinfo
  {journal} {Optica}\ }\textbf {\bibinfo {volume} {8}},\ \bibinfo {pages}
  {1243} (\bibinfo {year} {2021})}\BibitemShut {NoStop}%
\bibitem [{\citenamefont {Ayuso}\ \emph {et~al.}(2022)\citenamefont {Ayuso},
  \citenamefont {Ordonez}, \citenamefont {Decleva}, \citenamefont {Ivanov},\
  and\ \citenamefont {Smirnova}}]{Ayuso2022}%
  \BibitemOpen
  \bibfield  {author} {\bibinfo {author} {\bibfnamefont {D.}~\bibnamefont
  {Ayuso}}, \bibinfo {author} {\bibfnamefont {A.}~\bibnamefont {Ordonez},
  \bibfnamefont {F.}}, \bibinfo {author} {\bibfnamefont {P.}~\bibnamefont
  {Decleva}}, \bibinfo {author} {\bibfnamefont {M.}~\bibnamefont {Ivanov}},\
  and\ \bibinfo {author} {\bibfnamefont {O.}~\bibnamefont {Smirnova}},\
  }\bibfield  {title} {\bibinfo {title} {Strong chiral response in
  non-collinear high harmonic generation driven by purely electric-dipole
  interactions.},\ }\href@noop {} {\bibfield  {journal} {\bibinfo  {journal}
  {Opt. Express}\ }\textbf {\bibinfo {volume} {30}},\ \bibinfo {pages} {4659}
  (\bibinfo {year} {2022})}\BibitemShut {NoStop}%
\bibitem [{\citenamefont {St\"{o}hr}\ \emph {et~al.}(1998)\citenamefont
  {St\"{o}hr}, \citenamefont {Padmore}, \citenamefont {Anders}, \citenamefont
  {Stammler},\ and\ \citenamefont {Scheinfein}}]{sto1998}%
  \BibitemOpen
  \bibfield  {author} {\bibinfo {author} {\bibfnamefont {J.}~\bibnamefont
  {St\"{o}hr}}, \bibinfo {author} {\bibfnamefont {H.~A.}\ \bibnamefont
  {Padmore}}, \bibinfo {author} {\bibfnamefont {S.}~\bibnamefont {Anders}},
  \bibinfo {author} {\bibfnamefont {T.}~\bibnamefont {Stammler}},\ and\
  \bibinfo {author} {\bibfnamefont {M.~R.}\ \bibnamefont {Scheinfein}},\
  }\bibfield  {title} {\bibinfo {title} {Principles of x-ray magnetic dichroism
  spectromicroscopy},\ }\href@noop {} {\bibfield  {journal} {\bibinfo
  {journal} {Surf. Rev. Lett.}\ }\textbf {\bibinfo {volume} {5}},\ \bibinfo
  {pages} {1297} (\bibinfo {year} {1998})}\BibitemShut {NoStop}%
\bibitem [{\citenamefont {Sarkar}\ \emph {et~al.}(2019)\citenamefont {Sarkar},
  \citenamefont {Behunin},\ and\ \citenamefont {Gibbs}}]{sar2019}%
  \BibitemOpen
  \bibfield  {author} {\bibinfo {author} {\bibfnamefont {S.}~\bibnamefont
  {Sarkar}}, \bibinfo {author} {\bibfnamefont {R.~O.}\ \bibnamefont
  {Behunin}},\ and\ \bibinfo {author} {\bibfnamefont {J.~G.}\ \bibnamefont
  {Gibbs}},\ }\bibfield  {title} {\bibinfo {title} {Shape-dependent,
  chiro-optical response of uv-active, nanohelix metamaterials},\ }\href@noop
  {} {\bibfield  {journal} {\bibinfo  {journal} {Nano Lett.}\ }\textbf
  {\bibinfo {volume} {19}},\ \bibinfo {pages} {8089} (\bibinfo {year}
  {2019})}\BibitemShut {NoStop}%
\bibitem [{\citenamefont {Verbeeck}\ \emph {et~al.}(2018)\citenamefont
  {Verbeeck}, \citenamefont {Béché}, \citenamefont {Müller-Caspary},
  \citenamefont {Guzzinati}, \citenamefont {Luong},\ and\ \citenamefont {{Den
  Hertog}}}]{ver2018}%
  \BibitemOpen
  \bibfield  {author} {\bibinfo {author} {\bibfnamefont {J.}~\bibnamefont
  {Verbeeck}}, \bibinfo {author} {\bibfnamefont {A.}~\bibnamefont {Béché}},
  \bibinfo {author} {\bibfnamefont {K.}~\bibnamefont {Müller-Caspary}},
  \bibinfo {author} {\bibfnamefont {G.}~\bibnamefont {Guzzinati}}, \bibinfo
  {author} {\bibfnamefont {M.~A.}\ \bibnamefont {Luong}},\ and\ \bibinfo
  {author} {\bibfnamefont {M.}~\bibnamefont {{Den Hertog}}},\ }\bibfield
  {title} {\bibinfo {title} {Demonstration of a 2×2 programmable phase plate
  for electrons},\ }\href
  {https://doi.org/https://doi.org/10.1016/j.ultramic.2018.03.017} {\bibfield
  {journal} {\bibinfo  {journal} {Ultramicroscopy}\ }\textbf {\bibinfo {volume}
  {190}},\ \bibinfo {pages} {58} (\bibinfo {year} {2018})}\BibitemShut
  {NoStop}%
\bibitem [{\citenamefont {Pozzi}\ \emph {et~al.}(2020)\citenamefont {Pozzi},
  \citenamefont {Grillo}, \citenamefont {Lu}, \citenamefont {Tavabi},
  \citenamefont {Karimi},\ and\ \citenamefont {Dunin-Borkowski}}]{poz2020}%
  \BibitemOpen
  \bibfield  {author} {\bibinfo {author} {\bibfnamefont {G.}~\bibnamefont
  {Pozzi}}, \bibinfo {author} {\bibfnamefont {V.}~\bibnamefont {Grillo}},
  \bibinfo {author} {\bibfnamefont {P.-H.}\ \bibnamefont {Lu}}, \bibinfo
  {author} {\bibfnamefont {A.~H.}\ \bibnamefont {Tavabi}}, \bibinfo {author}
  {\bibfnamefont {E.}~\bibnamefont {Karimi}},\ and\ \bibinfo {author}
  {\bibfnamefont {R.~E.}\ \bibnamefont {Dunin-Borkowski}},\ }\bibfield  {title}
  {\bibinfo {title} {Design of electrostatic phase elements for sorting the
  orbital angular momentum of electrons},\ }\href
  {https://doi.org/https://doi.org/10.1016/j.ultramic.2019.112861} {\bibfield
  {journal} {\bibinfo  {journal} {Ultramicroscopy}\ }\textbf {\bibinfo {volume}
  {208}},\ \bibinfo {pages} {112861} (\bibinfo {year} {2020})}\BibitemShut
  {NoStop}%
\bibitem [{\citenamefont {Béché}\ \emph {et~al.}(2014)\citenamefont
  {Béché}, \citenamefont {Van~Boxem},\ and\ \citenamefont
  {Van~Tendeloo}}]{beh2014}%
  \BibitemOpen
  \bibfield  {author} {\bibinfo {author} {\bibfnamefont {A.}~\bibnamefont
  {Béché}}, \bibinfo {author} {\bibfnamefont {R.}~\bibnamefont {Van~Boxem}},\
  and\ \bibinfo {author} {\bibfnamefont {G.~e.~a.}\ \bibnamefont
  {Van~Tendeloo}},\ }\bibfield  {title} {\bibinfo {title} {Magnetic monopole
  field exposed by electrons.},\ }\href@noop {} {\bibfield  {journal} {\bibinfo
   {journal} {Nat. Phys.}\ }\textbf {\bibinfo {volume} {10}},\ \bibinfo {pages}
  {26–29} (\bibinfo {year} {2014})}\BibitemShut {NoStop}%
\bibitem [{\citenamefont {Voloch-Bloch}\ \emph {et~al.}(2013)\citenamefont
  {Voloch-Bloch}, \citenamefont {Lereah}, \citenamefont {Lilach}, \citenamefont
  {Gover},\ and\ \citenamefont {Arie}}]{blo2013}%
  \BibitemOpen
  \bibfield  {author} {\bibinfo {author} {\bibfnamefont {N.}~\bibnamefont
  {Voloch-Bloch}}, \bibinfo {author} {\bibfnamefont {Y.}~\bibnamefont
  {Lereah}}, \bibinfo {author} {\bibfnamefont {Y.}~\bibnamefont {Lilach}},
  \bibinfo {author} {\bibfnamefont {A.}~\bibnamefont {Gover}},\ and\ \bibinfo
  {author} {\bibfnamefont {A.}~\bibnamefont {Arie}},\ }\bibfield  {title}
  {\bibinfo {title} {Generation of electron airy beams},\ }\href
  {https://doi.org/10.1038/nature11840} {\bibfield  {journal} {\bibinfo
  {journal} {Nature}\ }\textbf {\bibinfo {volume} {494}},\ \bibinfo {pages}
  {331} (\bibinfo {year} {2013})}\BibitemShut {NoStop}%
\bibitem [{\citenamefont {Grillo}\ \emph {et~al.}(2014)\citenamefont {Grillo},
  \citenamefont {Karimi}, \citenamefont {Gazzadi}, \citenamefont {Frabboni},
  \citenamefont {Dennis},\ and\ \citenamefont {Boyd}}]{gri2014}%
  \BibitemOpen
  \bibfield  {author} {\bibinfo {author} {\bibfnamefont {V.}~\bibnamefont
  {Grillo}}, \bibinfo {author} {\bibfnamefont {E.}~\bibnamefont {Karimi}},
  \bibinfo {author} {\bibfnamefont {G.~C.}\ \bibnamefont {Gazzadi}}, \bibinfo
  {author} {\bibfnamefont {S.}~\bibnamefont {Frabboni}}, \bibinfo {author}
  {\bibfnamefont {M.~R.}\ \bibnamefont {Dennis}},\ and\ \bibinfo {author}
  {\bibfnamefont {R.~W.}\ \bibnamefont {Boyd}},\ }\bibfield  {title} {\bibinfo
  {title} {Generation of nondiffracting electron bessel beams},\ }\href
  {https://doi.org/10.1103/PhysRevX.4.011013} {\bibfield  {journal} {\bibinfo
  {journal} {Phys. Rev. X}\ }\textbf {\bibinfo {volume} {4}},\ \bibinfo {pages}
  {011013} (\bibinfo {year} {2014})}\BibitemShut {NoStop}%
\bibitem [{\citenamefont {Schwartz}\ \emph {et~al.}(2019)\citenamefont
  {Schwartz}, \citenamefont {Axelrod}, \citenamefont {Campbell}, \citenamefont
  {Turnbaugh}, \citenamefont {Glaeser},\ and\ \citenamefont
  {Müller}}]{sch2019}%
  \BibitemOpen
  \bibfield  {author} {\bibinfo {author} {\bibfnamefont {O.}~\bibnamefont
  {Schwartz}}, \bibinfo {author} {\bibfnamefont {J.}~\bibnamefont {Axelrod}},
  \bibinfo {author} {\bibfnamefont {S.}~\bibnamefont {Campbell}}, \bibinfo
  {author} {\bibfnamefont {C.}~\bibnamefont {Turnbaugh}}, \bibinfo {author}
  {\bibfnamefont {R.}~\bibnamefont {Glaeser}},\ and\ \bibinfo {author}
  {\bibfnamefont {H.}~\bibnamefont {Müller}},\ }\bibfield  {title} {\bibinfo
  {title} {Laser phase plate for transmission electron microscopy},\ }\href
  {https://doi.org/10.1038/s41592-019-0552-2} {\bibfield  {journal} {\bibinfo
  {journal} {Nat. Methods}\ }\textbf {\bibinfo {volume} {16}},\ \bibinfo
  {pages} {1016} (\bibinfo {year} {2019})}\BibitemShut {NoStop}%
\bibitem [{\citenamefont {Feist}\ \emph {et~al.}(2015)\citenamefont {Feist},
  \citenamefont {Priebe}, \citenamefont {Schauss}, \citenamefont {Yalunin},
  \citenamefont {Schäfer},\ and\ \citenamefont {Ropers}}]{fei2015}%
  \BibitemOpen
  \bibfield  {author} {\bibinfo {author} {\bibfnamefont {A.}~\bibnamefont
  {Feist}}, \bibinfo {author} {\bibfnamefont {K.}~\bibnamefont {Priebe}},
  \bibinfo {author} {\bibfnamefont {J.}~\bibnamefont {Schauss}}, \bibinfo
  {author} {\bibfnamefont {S.}~\bibnamefont {Yalunin}}, \bibinfo {author}
  {\bibfnamefont {S.}~\bibnamefont {Schäfer}},\ and\ \bibinfo {author}
  {\bibfnamefont {C.}~\bibnamefont {Ropers}},\ }\bibfield  {title} {\bibinfo
  {title} {Quantum coherent optical phase modulation in an ultrafast
  transmission electron microscope},\ }\href
  {https://doi.org/10.1038/nature14463} {\bibfield  {journal} {\bibinfo
  {journal} {Nature}\ }\textbf {\bibinfo {volume} {521}},\ \bibinfo {pages}
  {200} (\bibinfo {year} {2015})}\BibitemShut {NoStop}%
\bibitem [{\citenamefont {Vanacore}\ \emph {et~al.}(2018)\citenamefont
  {Vanacore}, \citenamefont {Madan}, \citenamefont {Berruto}, \citenamefont
  {Wang}, \citenamefont {Pomarico}, \citenamefont {Lamb}, \citenamefont
  {McGrouther}, \citenamefont {Kaminer}, \citenamefont {Barwick},\ and\
  \citenamefont {Carbone}}]{van2018}%
  \BibitemOpen
  \bibfield  {author} {\bibinfo {author} {\bibfnamefont {G.~M.}\ \bibnamefont
  {Vanacore}}, \bibinfo {author} {\bibfnamefont {I.}~\bibnamefont {Madan}},
  \bibinfo {author} {\bibfnamefont {G.}~\bibnamefont {Berruto}}, \bibinfo
  {author} {\bibfnamefont {K.}~\bibnamefont {Wang}}, \bibinfo {author}
  {\bibfnamefont {E.}~\bibnamefont {Pomarico}}, \bibinfo {author}
  {\bibfnamefont {R.~J.}\ \bibnamefont {Lamb}}, \bibinfo {author}
  {\bibfnamefont {D.}~\bibnamefont {McGrouther}}, \bibinfo {author}
  {\bibfnamefont {I.}~\bibnamefont {Kaminer}}, \bibinfo {author} {\bibfnamefont
  {F.~J.}\ \bibnamefont {Barwick}, \bibfnamefont {B.and García de~Abajo}},\
  and\ \bibinfo {author} {\bibfnamefont {F.}~\bibnamefont {Carbone}},\
  }\bibfield  {title} {\bibinfo {title} {Attosecond coherent control of
  free-electron wave functions using semi-infinite light fields.},\ }\href@noop
  {} {\bibfield  {journal} {\bibinfo  {journal} {Nat. Commun.}\ }\textbf
  {\bibinfo {volume} {9}},\ \bibinfo {pages} {2694} (\bibinfo {year}
  {2018})}\BibitemShut {NoStop}%
\bibitem [{\citenamefont {Konečná}\ and\ \citenamefont {Garcia~de
  Abajo}(2020)}]{kon2020}%
  \BibitemOpen
  \bibfield  {author} {\bibinfo {author} {\bibfnamefont {A.}~\bibnamefont
  {Konečná}}\ and\ \bibinfo {author} {\bibfnamefont {J.}~\bibnamefont
  {Garcia~de Abajo}},\ }\bibfield  {title} {\bibinfo {title} {Electron beam
  aberration correction using optical near fields},\ }\href
  {https://doi.org/10.1103/PhysRevLett.125.030801} {\bibfield  {journal}
  {\bibinfo  {journal} {Phys. Rev. Lett.}\ }\textbf {\bibinfo {volume} {125}}
  (\bibinfo {year} {2020})}\BibitemShut {NoStop}%
\bibitem [{\citenamefont {Feist}\ \emph {et~al.}(2020)\citenamefont {Feist},
  \citenamefont {Yalunin}, \citenamefont {Sch\"afer},\ and\ \citenamefont
  {Ropers}}]{fei2020}%
  \BibitemOpen
  \bibfield  {author} {\bibinfo {author} {\bibfnamefont {A.}~\bibnamefont
  {Feist}}, \bibinfo {author} {\bibfnamefont {S.~V.}\ \bibnamefont {Yalunin}},
  \bibinfo {author} {\bibfnamefont {S.}~\bibnamefont {Sch\"afer}},\ and\
  \bibinfo {author} {\bibfnamefont {C.}~\bibnamefont {Ropers}},\ }\bibfield
  {title} {\bibinfo {title} {High-purity free-electron momentum states prepared
  by three-dimensional optical phase modulation},\ }\href
  {https://doi.org/10.1103/PhysRevResearch.2.043227} {\bibfield  {journal}
  {\bibinfo  {journal} {Phys. Rev. Res.}\ }\textbf {\bibinfo {volume} {2}},\
  \bibinfo {pages} {043227} (\bibinfo {year} {2020})}\BibitemShut {NoStop}%
\bibitem [{\citenamefont {Garcia~de Abajo}\ and\ \citenamefont
  {Konečná}(2021)}]{aba2021}%
  \BibitemOpen
  \bibfield  {author} {\bibinfo {author} {\bibfnamefont {J.}~\bibnamefont
  {Garcia~de Abajo}}\ and\ \bibinfo {author} {\bibfnamefont {A.}~\bibnamefont
  {Konečná}},\ }\bibfield  {title} {\bibinfo {title} {Optical modulation of
  electron beams in free space},\ }\href
  {https://doi.org/10.1103/PhysRevLett.126.123901} {\bibfield  {journal}
  {\bibinfo  {journal} {Phys. Rev. Lett.}\ }\textbf {\bibinfo {volume} {126}}
  (\bibinfo {year} {2021})}\BibitemShut {NoStop}%
\bibitem [{\citenamefont {Chirita~Mihaila}\ \emph {et~al.}(2022)\citenamefont
  {Chirita~Mihaila}, \citenamefont {Weber}, \citenamefont {Schneller},
  \citenamefont {Grandits}, \citenamefont {Nimmrichter},\ and\ \citenamefont
  {Juffmann}}]{chi2022}%
  \BibitemOpen
  \bibfield  {author} {\bibinfo {author} {\bibfnamefont {M.~C.}\ \bibnamefont
  {Chirita~Mihaila}}, \bibinfo {author} {\bibfnamefont {P.}~\bibnamefont
  {Weber}}, \bibinfo {author} {\bibfnamefont {M.}~\bibnamefont {Schneller}},
  \bibinfo {author} {\bibfnamefont {L.}~\bibnamefont {Grandits}}, \bibinfo
  {author} {\bibfnamefont {S.}~\bibnamefont {Nimmrichter}},\ and\ \bibinfo
  {author} {\bibfnamefont {T.}~\bibnamefont {Juffmann}},\ }\bibfield  {title}
  {\bibinfo {title} {Transverse electron-beam shaping with light},\ }\href
  {https://doi.org/10.1103/PhysRevX.12.031043} {\bibfield  {journal} {\bibinfo
  {journal} {Phys. Rev. X}\ }\textbf {\bibinfo {volume} {12}},\ \bibinfo
  {pages} {031043} (\bibinfo {year} {2022})}\BibitemShut {NoStop}%
\bibitem [{\citenamefont {Verbeeck}\ \emph {et~al.}(2010)\citenamefont
  {Verbeeck}, \citenamefont {Tian},\ and\ \citenamefont
  {Schattschneider}}]{ver2010}%
  \BibitemOpen
  \bibfield  {author} {\bibinfo {author} {\bibfnamefont {J.}~\bibnamefont
  {Verbeeck}}, \bibinfo {author} {\bibfnamefont {H.}~\bibnamefont {Tian}},\
  and\ \bibinfo {author} {\bibfnamefont {P.}~\bibnamefont {Schattschneider}},\
  }\bibfield  {title} {\bibinfo {title} {Production and application of electron
  vortex beams},\ }\href {https://doi.org/10.1038/nature09366} {\bibfield
  {journal} {\bibinfo  {journal} {Nature}\ }\textbf {\bibinfo {volume} {467}},\
  \bibinfo {pages} {301} (\bibinfo {year} {2010})}\BibitemShut {NoStop}%
\bibitem [{\citenamefont {Bliokh}\ \emph {et~al.}(2017)\citenamefont {Bliokh},
  \citenamefont {Ivanov}, \citenamefont {Guzzinati}, \citenamefont {Clark},
  \citenamefont {{Van Boxem}}, \citenamefont {Béché}, \citenamefont
  {Juchtmans}, \citenamefont {Alonso}, \citenamefont {Schattschneider},
  \citenamefont {Nori},\ and\ \citenamefont {Verbeeck}}]{BLIOKH20171}%
  \BibitemOpen
  \bibfield  {author} {\bibinfo {author} {\bibfnamefont {K.}~\bibnamefont
  {Bliokh}}, \bibinfo {author} {\bibfnamefont {I.}~\bibnamefont {Ivanov}},
  \bibinfo {author} {\bibfnamefont {G.}~\bibnamefont {Guzzinati}}, \bibinfo
  {author} {\bibfnamefont {L.}~\bibnamefont {Clark}}, \bibinfo {author}
  {\bibfnamefont {R.}~\bibnamefont {{Van Boxem}}}, \bibinfo {author}
  {\bibfnamefont {A.}~\bibnamefont {Béché}}, \bibinfo {author} {\bibfnamefont
  {R.}~\bibnamefont {Juchtmans}}, \bibinfo {author} {\bibfnamefont
  {M.}~\bibnamefont {Alonso}}, \bibinfo {author} {\bibfnamefont
  {P.}~\bibnamefont {Schattschneider}}, \bibinfo {author} {\bibfnamefont
  {F.}~\bibnamefont {Nori}},\ and\ \bibinfo {author} {\bibfnamefont
  {J.}~\bibnamefont {Verbeeck}},\ }\bibfield  {title} {\bibinfo {title} {Theory
  and applications of free-electron vortex states},\ }\href
  {https://doi.org/https://doi.org/10.1016/j.physrep.2017.05.006} {\bibfield
  {journal} {\bibinfo  {journal} {Phys. Rep.-Rev. Sec. Phys. Lett.}\ }\textbf
  {\bibinfo {volume} {690}},\ \bibinfo {pages} {1} (\bibinfo {year}
  {2017})}\BibitemShut {NoStop}%
\bibitem [{\citenamefont {Uchida}\ and\ \citenamefont
  {Tonomura}(2010)}]{uch2010}%
  \BibitemOpen
  \bibfield  {author} {\bibinfo {author} {\bibfnamefont {M.}~\bibnamefont
  {Uchida}}\ and\ \bibinfo {author} {\bibfnamefont {A.}~\bibnamefont
  {Tonomura}},\ }\bibfield  {title} {\bibinfo {title} {Generation of electron
  beams carrying orbital angular momentum},\ }\href
  {https://doi.org/10.1038/nature08904} {\bibfield  {journal} {\bibinfo
  {journal} {Nature}\ }\textbf {\bibinfo {volume} {464}},\ \bibinfo {pages}
  {737} (\bibinfo {year} {2010})}\BibitemShut {NoStop}%
\bibitem [{\citenamefont {Tavabi}\ \emph {et~al.}(2022)\citenamefont {Tavabi},
  \citenamefont {Rosi}, \citenamefont {Roncaglia}, \citenamefont {Rotunno},
  \citenamefont {Beleggia}, \citenamefont {Lu}, \citenamefont {Belsito},
  \citenamefont {Pozzi}, \citenamefont {Frabboni}, \citenamefont {Tiemeijer},
  \citenamefont {Dunin-Borkowski},\ and\ \citenamefont {Grillo}}]{tav2022}%
  \BibitemOpen
  \bibfield  {author} {\bibinfo {author} {\bibfnamefont {A.~H.}\ \bibnamefont
  {Tavabi}}, \bibinfo {author} {\bibfnamefont {P.}~\bibnamefont {Rosi}},
  \bibinfo {author} {\bibfnamefont {A.}~\bibnamefont {Roncaglia}}, \bibinfo
  {author} {\bibfnamefont {E.}~\bibnamefont {Rotunno}}, \bibinfo {author}
  {\bibfnamefont {M.}~\bibnamefont {Beleggia}}, \bibinfo {author}
  {\bibfnamefont {P.-H.}\ \bibnamefont {Lu}}, \bibinfo {author} {\bibfnamefont
  {L.}~\bibnamefont {Belsito}}, \bibinfo {author} {\bibfnamefont
  {G.}~\bibnamefont {Pozzi}}, \bibinfo {author} {\bibfnamefont
  {S.}~\bibnamefont {Frabboni}}, \bibinfo {author} {\bibfnamefont
  {P.}~\bibnamefont {Tiemeijer}}, \bibinfo {author} {\bibfnamefont {R.~E.}\
  \bibnamefont {Dunin-Borkowski}},\ and\ \bibinfo {author} {\bibfnamefont
  {V.}~\bibnamefont {Grillo}},\ }\bibfield  {title} {\bibinfo {title}
  {{Generation of electron vortex beams with over 1000 orbital angular momentum
  quanta using a tunable electrostatic spiral phase plate}},\ }\href@noop {}
  {\bibfield  {journal} {\bibinfo  {journal} {Appl. Phys. Lett.}\ }\textbf
  {\bibinfo {volume} {121}},\ \bibinfo {pages} {073506} (\bibinfo {year}
  {2022})}\BibitemShut {NoStop}%
\bibitem [{\citenamefont {Yu}\ \emph {et~al.}(2023)\citenamefont {Yu},
  \citenamefont {Huo}, \citenamefont {Liu}, \citenamefont {Zhu}, \citenamefont
  {Agrawal}, \citenamefont {Lu},\ and\ \citenamefont {Xu}}]{yu2023}%
  \BibitemOpen
  \bibfield  {author} {\bibinfo {author} {\bibfnamefont {R.}~\bibnamefont
  {Yu}}, \bibinfo {author} {\bibfnamefont {P.}~\bibnamefont {Huo}}, \bibinfo
  {author} {\bibfnamefont {M.}~\bibnamefont {Liu}}, \bibinfo {author}
  {\bibfnamefont {W.}~\bibnamefont {Zhu}}, \bibinfo {author} {\bibfnamefont
  {A.}~\bibnamefont {Agrawal}}, \bibinfo {author} {\bibfnamefont {Y.-q.}\
  \bibnamefont {Lu}},\ and\ \bibinfo {author} {\bibfnamefont {T.}~\bibnamefont
  {Xu}},\ }\bibfield  {title} {\bibinfo {title} {Generation of perfect electron
  vortex beam with a customized beam size independent of orbital angular
  momentum},\ }\href@noop {} {\bibfield  {journal} {\bibinfo  {journal} {Nano
  Lett.}\ }\textbf {\bibinfo {volume} {23}},\ \bibinfo {pages} {2436} (\bibinfo
  {year} {2023})}\BibitemShut {NoStop}%
\bibitem [{\citenamefont {Vanacore}\ \emph {et~al.}(2019)\citenamefont
  {Vanacore}, \citenamefont {Berruto}, \citenamefont {Madan}, \citenamefont
  {Pomarico}, \citenamefont {Biagioni}, \citenamefont {Lamb}, \citenamefont
  {McGrouther}, \citenamefont {Reinhardt}, \citenamefont {Kaminer},
  \citenamefont {Barwick}, \citenamefont {Larocque}, \citenamefont {Grillo},
  \citenamefont {Karimi}, \citenamefont {Garcia~de Abajo},\ and\ \citenamefont
  {Carbone}}]{van2019}%
  \BibitemOpen
  \bibfield  {author} {\bibinfo {author} {\bibfnamefont {G.}~\bibnamefont
  {Vanacore}}, \bibinfo {author} {\bibfnamefont {G.}~\bibnamefont {Berruto}},
  \bibinfo {author} {\bibfnamefont {I.}~\bibnamefont {Madan}}, \bibinfo
  {author} {\bibfnamefont {E.}~\bibnamefont {Pomarico}}, \bibinfo {author}
  {\bibfnamefont {P.}~\bibnamefont {Biagioni}}, \bibinfo {author}
  {\bibfnamefont {R.}~\bibnamefont {Lamb}}, \bibinfo {author} {\bibfnamefont
  {D.}~\bibnamefont {McGrouther}}, \bibinfo {author} {\bibfnamefont
  {O.}~\bibnamefont {Reinhardt}}, \bibinfo {author} {\bibfnamefont
  {I.}~\bibnamefont {Kaminer}}, \bibinfo {author} {\bibfnamefont
  {B.}~\bibnamefont {Barwick}}, \bibinfo {author} {\bibfnamefont
  {H.}~\bibnamefont {Larocque}}, \bibinfo {author} {\bibfnamefont
  {V.}~\bibnamefont {Grillo}}, \bibinfo {author} {\bibfnamefont
  {E.}~\bibnamefont {Karimi}}, \bibinfo {author} {\bibfnamefont
  {J.}~\bibnamefont {Garcia~de Abajo}},\ and\ \bibinfo {author} {\bibfnamefont
  {F.}~\bibnamefont {Carbone}},\ }\bibfield  {title} {\bibinfo {title}
  {Ultrafast generation and control of an electron vortex beam via chiral
  plasmonic near fields},\ }\href {https://doi.org/10.1038/s41563-019-0336-1}
  {\bibfield  {journal} {\bibinfo  {journal} {Nat. Mater.}\ }\textbf {\bibinfo
  {volume} {18}},\ \bibinfo {pages} {573} (\bibinfo {year} {2019})}\BibitemShut
  {NoStop}%
\bibitem [{\citenamefont {Kozak}(2021)}]{koz2021}%
  \BibitemOpen
  \bibfield  {author} {\bibinfo {author} {\bibfnamefont {M.}~\bibnamefont
  {Kozak}},\ }\bibfield  {title} {\bibinfo {title} {Electron vortex beam
  generation via chiral light-induced inelastic ponderomotive scattering},\
  }\href {https://doi.org/10.1021/acsphotonics.0c01650} {\bibfield  {journal}
  {\bibinfo  {journal} {ACS Photonics}\ }\textbf {\bibinfo {volume} {8}},\
  \bibinfo {pages} {431–435} (\bibinfo {year} {2021})}\BibitemShut {NoStop}%
\bibitem [{\citenamefont {Zanfrognini}\ \emph {et~al.}(2019)\citenamefont
  {Zanfrognini}, \citenamefont {Rotunno}, \citenamefont {Frabboni},
  \citenamefont {Sit}, \citenamefont {Karimi}, \citenamefont {Hohenester},\
  and\ \citenamefont {Grillo}}]{zan2019}%
  \BibitemOpen
  \bibfield  {author} {\bibinfo {author} {\bibfnamefont {M.}~\bibnamefont
  {Zanfrognini}}, \bibinfo {author} {\bibfnamefont {E.}~\bibnamefont
  {Rotunno}}, \bibinfo {author} {\bibfnamefont {S.}~\bibnamefont {Frabboni}},
  \bibinfo {author} {\bibfnamefont {A.}~\bibnamefont {Sit}}, \bibinfo {author}
  {\bibfnamefont {E.}~\bibnamefont {Karimi}}, \bibinfo {author} {\bibfnamefont
  {U.}~\bibnamefont {Hohenester}},\ and\ \bibinfo {author} {\bibfnamefont
  {V.}~\bibnamefont {Grillo}},\ }\bibfield  {title} {\bibinfo {title} {Orbital
  angular momentum and energy loss characterization of plasmonic excitations in
  metallic nanostructures in tem},\ }\href
  {https://doi.org/10.1021/acsphotonics.9b00131} {\bibfield  {journal}
  {\bibinfo  {journal} {ACS Photonics}\ }\textbf {\bibinfo {volume} {6}},\
  \bibinfo {pages} {620} (\bibinfo {year} {2019})}\BibitemShut {NoStop}%
\bibitem [{\citenamefont {Guzzinati}\ \emph {et~al.}(2017)\citenamefont
  {Guzzinati}, \citenamefont {Béché}, \citenamefont {Lourenço-Martins},
  \citenamefont {Martin}, \citenamefont {Kociak},\ and\ \citenamefont
  {Verbeeck}}]{guz2017}%
  \BibitemOpen
  \bibfield  {author} {\bibinfo {author} {\bibfnamefont {G.}~\bibnamefont
  {Guzzinati}}, \bibinfo {author} {\bibfnamefont {A.}~\bibnamefont {Béché}},
  \bibinfo {author} {\bibfnamefont {H.}~\bibnamefont {Lourenço-Martins}},
  \bibinfo {author} {\bibfnamefont {J.}~\bibnamefont {Martin}}, \bibinfo
  {author} {\bibfnamefont {M.}~\bibnamefont {Kociak}},\ and\ \bibinfo {author}
  {\bibfnamefont {J.}~\bibnamefont {Verbeeck}},\ }\bibfield  {title} {\bibinfo
  {title} {Probing the symmetry of the potential of localized surface plasmon
  resonances with phase-shaped electron beams},\ }\href
  {https://doi.org/10.1038/ncomms14999} {\bibfield  {journal} {\bibinfo
  {journal} {Nat. Commun.}\ }\textbf {\bibinfo {volume} {8}},\ \bibinfo {pages}
  {14999} (\bibinfo {year} {2017})}\BibitemShut {NoStop}%
\bibitem [{\citenamefont {Asenjo-Garcia}\ and\ \citenamefont {Garcia~de
  Abajo}(2014)}]{ase2014}%
  \BibitemOpen
  \bibfield  {author} {\bibinfo {author} {\bibfnamefont {A.}~\bibnamefont
  {Asenjo-Garcia}}\ and\ \bibinfo {author} {\bibfnamefont {J.}~\bibnamefont
  {Garcia~de Abajo}},\ }\bibfield  {title} {\bibinfo {title} {Dichroism in the
  interaction between vortex electron beams, plasmons, and molecules},\ }\href
  {https://doi.org/10.1103/PhysRevLett.113.066102} {\bibfield  {journal}
  {\bibinfo  {journal} {Phys. Rev. Lett.}\ }\textbf {\bibinfo {volume} {113}},\
  \bibinfo {pages} {066102} (\bibinfo {year} {2014})}\BibitemShut {NoStop}%
\bibitem [{\citenamefont {Lourenço-Martins}\ \emph {et~al.}(2021)\citenamefont
  {Lourenço-Martins}, \citenamefont {Gérard},\ and\ \citenamefont
  {Kociak}}]{lou2021}%
  \BibitemOpen
  \bibfield  {author} {\bibinfo {author} {\bibfnamefont {H.}~\bibnamefont
  {Lourenço-Martins}}, \bibinfo {author} {\bibfnamefont {D.}~\bibnamefont
  {Gérard}},\ and\ \bibinfo {author} {\bibfnamefont {M.}~\bibnamefont
  {Kociak}},\ }\bibfield  {title} {\bibinfo {title} {Optical polarization
  analogue in free electron beams},\ }\href
  {https://doi.org/10.1038/s41567-021-01163-w} {\bibfield  {journal} {\bibinfo
  {journal} {Nat. Phys.}\ }\textbf {\bibinfo {volume} {17}},\ \bibinfo {pages}
  {598} (\bibinfo {year} {2021})}\BibitemShut {NoStop}%
\bibitem [{\citenamefont {Kone\ifmmode~\check{c}\else \v{c}\fi{}n\'a}\ \emph
  {et~al.}(2023)\citenamefont {Kone\ifmmode~\check{c}\else \v{c}\fi{}n\'a},
  \citenamefont {Schmidt}, \citenamefont {Hillenbrand},\ and\ \citenamefont
  {Aizpurua}}]{kon2023}%
  \BibitemOpen
  \bibfield  {author} {\bibinfo {author} {\bibfnamefont {A.}~\bibnamefont
  {Kone\ifmmode~\check{c}\else \v{c}\fi{}n\'a}}, \bibinfo {author}
  {\bibfnamefont {M.~K.}\ \bibnamefont {Schmidt}}, \bibinfo {author}
  {\bibfnamefont {R.}~\bibnamefont {Hillenbrand}},\ and\ \bibinfo {author}
  {\bibfnamefont {J.}~\bibnamefont {Aizpurua}},\ }\bibfield  {title} {\bibinfo
  {title} {Probing the electromagnetic response of dielectric antennas by
  vortex electron beams},\ }\href
  {https://doi.org/10.1103/PhysRevResearch.5.023192} {\bibfield  {journal}
  {\bibinfo  {journal} {Phys. Rev. Res.}\ }\textbf {\bibinfo {volume} {5}},\
  \bibinfo {pages} {023192} (\bibinfo {year} {2023})}\BibitemShut {NoStop}%
\bibitem [{\citenamefont {Harvey}\ \emph {et~al.}(2020)\citenamefont {Harvey},
  \citenamefont {Henke}, \citenamefont {Kfir}, \citenamefont
  {Lourenço-Martins}, \citenamefont {Feist}, \citenamefont {García~de
  Abajo},\ and\ \citenamefont {Ropers}}]{Har2020}%
  \BibitemOpen
  \bibfield  {author} {\bibinfo {author} {\bibfnamefont {T.~R.}\ \bibnamefont
  {Harvey}}, \bibinfo {author} {\bibfnamefont {J.-W.}\ \bibnamefont {Henke}},
  \bibinfo {author} {\bibfnamefont {O.}~\bibnamefont {Kfir}}, \bibinfo {author}
  {\bibfnamefont {H.}~\bibnamefont {Lourenço-Martins}}, \bibinfo {author}
  {\bibfnamefont {A.}~\bibnamefont {Feist}}, \bibinfo {author} {\bibfnamefont
  {F.~J.}\ \bibnamefont {García~de Abajo}},\ and\ \bibinfo {author}
  {\bibfnamefont {C.}~\bibnamefont {Ropers}},\ }\bibfield  {title} {\bibinfo
  {title} {Probing chirality with inelastic electron-light scattering},\
  }\href@noop {} {\bibfield  {journal} {\bibinfo  {journal} {Nano Lett.}\
  }\textbf {\bibinfo {volume} {20}},\ \bibinfo {pages} {4377} (\bibinfo {year}
  {2020})}\BibitemShut {NoStop}%
\bibitem [{\citenamefont {Bliokh}\ \emph {et~al.}(2015)\citenamefont {Bliokh},
  \citenamefont {Rodríguez-Fortuño}, \citenamefont {Nori},\ and\
  \citenamefont {Zayats}}]{bli2015}%
  \BibitemOpen
  \bibfield  {author} {\bibinfo {author} {\bibfnamefont {K.}~\bibnamefont
  {Bliokh}}, \bibinfo {author} {\bibfnamefont {F.}~\bibnamefont
  {Rodríguez-Fortuño}}, \bibinfo {author} {\bibfnamefont {F.}~\bibnamefont
  {Nori}},\ and\ \bibinfo {author} {\bibfnamefont {A.}~\bibnamefont {Zayats}},\
  }\bibfield  {title} {\bibinfo {title} {Spin-orbit interactions of light},\
  }\href {https://doi.org/10.1038/nphoton.2015.201} {\bibfield  {journal}
  {\bibinfo  {journal} {Nat. Photonics}\ }\textbf {\bibinfo {volume} {9}}
  (\bibinfo {year} {2015})}\BibitemShut {NoStop}%
\bibitem [{\citenamefont {Giulio}\ and\ \citenamefont
  {de~Abajo}(2020)}]{dig2020}%
  \BibitemOpen
  \bibfield  {author} {\bibinfo {author} {\bibfnamefont {V.~D.}\ \bibnamefont
  {Giulio}}\ and\ \bibinfo {author} {\bibfnamefont {F.~J.~G.}\ \bibnamefont
  {de~Abajo}},\ }\bibfield  {title} {\bibinfo {title} {Free-electron shaping
  using quantum light},\ }\href {https://doi.org/10.1364/optica.404598}
  {\bibfield  {journal} {\bibinfo  {journal} {Optica}\ }\textbf {\bibinfo
  {volume} {7}},\ \bibinfo {pages} {1820} (\bibinfo {year} {2020})}\BibitemShut
  {NoStop}%
\bibitem [{\citenamefont {Park}\ \emph {et~al.}(2010)\citenamefont {Park},
  \citenamefont {Lin},\ and\ \citenamefont {Zewail}}]{par2010}%
  \BibitemOpen
  \bibfield  {author} {\bibinfo {author} {\bibfnamefont {S.~T.}\ \bibnamefont
  {Park}}, \bibinfo {author} {\bibfnamefont {M.}~\bibnamefont {Lin}},\ and\
  \bibinfo {author} {\bibfnamefont {A.~H.}\ \bibnamefont {Zewail}},\ }\bibfield
   {title} {\bibinfo {title} {Photon-induced near-field electron microscopy
  (pinem): theoretical and experimental},\ }\href
  {https://doi.org/10.1088/1367-2630/12/12/123028} {\bibfield  {journal}
  {\bibinfo  {journal} {New J. Phys.}\ }\textbf {\bibinfo {volume} {12}},\
  \bibinfo {pages} {123028} (\bibinfo {year} {2010})}\BibitemShut {NoStop}%
\bibitem [{\citenamefont {Koz\'ak}\ \emph
  {et~al.}(2018{\natexlab{a}})\citenamefont {Koz\'ak}, \citenamefont
  {Eckstein}, \citenamefont {Schoenenberger},\ and\ \citenamefont
  {Hommelhoff}}]{koz2018}%
  \BibitemOpen
  \bibfield  {author} {\bibinfo {author} {\bibfnamefont {M.}~\bibnamefont
  {Koz\'ak}}, \bibinfo {author} {\bibfnamefont {T.}~\bibnamefont {Eckstein}},
  \bibinfo {author} {\bibfnamefont {N.}~\bibnamefont {Schoenenberger}},\ and\
  \bibinfo {author} {\bibfnamefont {P.}~\bibnamefont {Hommelhoff}},\ }\bibfield
   {title} {\bibinfo {title} {Inelastic ponderomotive scattering of electrons
  at a high-intensity optical travelling wave in vacuum},\ }\href
  {https://doi.org/10.1038/nphys4282} {\bibfield  {journal} {\bibinfo
  {journal} {Nat. Phys.}\ }\textbf {\bibinfo {volume} {14}} (\bibinfo {year}
  {2018}{\natexlab{a}})}\BibitemShut {NoStop}%
\bibitem [{\citenamefont {Koz\'ak}\ \emph
  {et~al.}(2018{\natexlab{b}})\citenamefont {Koz\'ak}, \citenamefont
  {Sch\"onenberger},\ and\ \citenamefont {Hommelhoff}}]{kozak2018}%
  \BibitemOpen
  \bibfield  {author} {\bibinfo {author} {\bibfnamefont {M.}~\bibnamefont
  {Koz\'ak}}, \bibinfo {author} {\bibfnamefont {N.}~\bibnamefont
  {Sch\"onenberger}},\ and\ \bibinfo {author} {\bibfnamefont {P.}~\bibnamefont
  {Hommelhoff}},\ }\bibfield  {title} {\bibinfo {title} {Ponderomotive
  generation and detection of attosecond free-electron pulse trains},\ }\href
  {https://doi.org/10.1103/PhysRevLett.120.103203} {\bibfield  {journal}
  {\bibinfo  {journal} {Phys. Rev. Lett.}\ }\textbf {\bibinfo {volume} {120}},\
  \bibinfo {pages} {103203} (\bibinfo {year} {2018}{\natexlab{b}})}\BibitemShut
  {NoStop}%
\bibitem [{\citenamefont {Smorenburg}\ \emph {et~al.}(2011)\citenamefont
  {Smorenburg}, \citenamefont {Kanters}, \citenamefont {Lassise}, \citenamefont
  {Brussaard}, \citenamefont {Kamp},\ and\ \citenamefont
  {Luiten}}]{PhysRevA.83.063810}%
  \BibitemOpen
  \bibfield  {author} {\bibinfo {author} {\bibfnamefont {P.~W.}\ \bibnamefont
  {Smorenburg}}, \bibinfo {author} {\bibfnamefont {J.~H.~M.}\ \bibnamefont
  {Kanters}}, \bibinfo {author} {\bibfnamefont {A.}~\bibnamefont {Lassise}},
  \bibinfo {author} {\bibfnamefont {G.~J.~H.}\ \bibnamefont {Brussaard}},
  \bibinfo {author} {\bibfnamefont {L.~P.~J.}\ \bibnamefont {Kamp}},\ and\
  \bibinfo {author} {\bibfnamefont {O.~J.}\ \bibnamefont {Luiten}},\ }\bibfield
   {title} {\bibinfo {title} {Polarization-dependent ponderomotive gradient
  force in a standing wave},\ }\href
  {https://doi.org/10.1103/PhysRevA.83.063810} {\bibfield  {journal} {\bibinfo
  {journal} {Phys. Rev. A}\ }\textbf {\bibinfo {volume} {83}},\ \bibinfo
  {pages} {063810} (\bibinfo {year} {2011})}\BibitemShut {NoStop}%
\bibitem [{\citenamefont {Axelrod}\ \emph {et~al.}(2020)\citenamefont
  {Axelrod}, \citenamefont {Campbell}, \citenamefont {Schwartz}, \citenamefont
  {Turnbaugh}, \citenamefont {Glaeser},\ and\ \citenamefont
  {M\"uller}}]{axe2020}%
  \BibitemOpen
  \bibfield  {author} {\bibinfo {author} {\bibfnamefont {J.~J.}\ \bibnamefont
  {Axelrod}}, \bibinfo {author} {\bibfnamefont {S.~L.}\ \bibnamefont
  {Campbell}}, \bibinfo {author} {\bibfnamefont {O.}~\bibnamefont {Schwartz}},
  \bibinfo {author} {\bibfnamefont {C.}~\bibnamefont {Turnbaugh}}, \bibinfo
  {author} {\bibfnamefont {R.~M.}\ \bibnamefont {Glaeser}},\ and\ \bibinfo
  {author} {\bibfnamefont {H.}~\bibnamefont {M\"uller}},\ }\bibfield  {title}
  {\bibinfo {title} {Observation of the relativistic reversal of the
  ponderomotive potential},\ }\href
  {https://doi.org/10.1103/PhysRevLett.124.174801} {\bibfield  {journal}
  {\bibinfo  {journal} {Phys. Rev. Lett.}\ }\textbf {\bibinfo {volume} {124}},\
  \bibinfo {pages} {174801} (\bibinfo {year} {2020})}\BibitemShut {NoStop}%
\bibitem [{\citenamefont {Schulman}(2012)}]{sch1981}%
  \BibitemOpen
  \bibfield  {author} {\bibinfo {author} {\bibfnamefont {L.}~\bibnamefont
  {Schulman}},\ }\href {https://books.google.cz/books?id=Ps0DcDKAEmIC} {\emph
  {\bibinfo {title} {Techniques and Applications of Path Integration}}},\ Dover
  Books on Physics\ (\bibinfo  {publisher} {Dover Publications},\ \bibinfo
  {year} {2012})\BibitemShut {NoStop}%
\bibitem [{\citenamefont {Tsarev}\ \emph {et~al.}(2023)\citenamefont {Tsarev},
  \citenamefont {Thurner},\ and\ \citenamefont {Baum}}]{tsa2023}%
  \BibitemOpen
  \bibfield  {author} {\bibinfo {author} {\bibfnamefont {M.}~\bibnamefont
  {Tsarev}}, \bibinfo {author} {\bibfnamefont {J.}~\bibnamefont {Thurner}},\
  and\ \bibinfo {author} {\bibfnamefont {P.}~\bibnamefont {Baum}},\ }\bibfield
  {title} {\bibinfo {title} {Nonlinear-optical quantum control of free-electron
  matter waves},\ }\href {https://doi.org/10.1038/s41567-023-02092-6}
  {\bibfield  {journal} {\bibinfo  {journal} {Nat. Phys.}\ }\textbf {\bibinfo
  {volume} {19}},\ \bibinfo {pages} {1350–1354} (\bibinfo {year}
  {2023})}\BibitemShut {NoStop}%
\bibitem [{SM()}]{SM}%
  \BibitemOpen
  \href@noop {} {}\bibinfo {note} {See Supplemental Material at
  \url{http://link.aps.org/supplemental/10.1103/PhysRevApplied.22.054017} for
  semiclassical model of the dichroic response.}\BibitemShut {Stop}%
\bibitem [{\citenamefont {Bohren}\ and\ \citenamefont
  {Huffman}(1998)}]{scattering1998}%
  \BibitemOpen
  \bibfield  {author} {\bibinfo {author} {\bibfnamefont {C.~F.}\ \bibnamefont
  {Bohren}}\ and\ \bibinfo {author} {\bibfnamefont {D.~R.}\ \bibnamefont
  {Huffman}},\ }\bibinfo {title} {Absorption and scattering by a sphere},\ in\
  \href {https://doi.org/https://doi.org/10.1002/9783527618156.ch4} {\emph
  {\bibinfo {booktitle} {Absorption and Scattering of Light by Small
  Particles}}}\ (\bibinfo  {publisher} {John Wiley \& Sons, Ltd},\ \bibinfo
  {year} {1998})\ Chap.~\bibinfo {chapter} {4}, pp.\ \bibinfo {pages}
  {82--129}\BibitemShut {NoStop}%
\bibitem [{\citenamefont {Raki\'{c}}\ \emph {et~al.}(1998)\citenamefont
  {Raki\'{c}}, \citenamefont {Djuri\v{s}i\'{c}}, \citenamefont {Elazar},\ and\
  \citenamefont {Majewski}}]{Rakic:98}%
  \BibitemOpen
  \bibfield  {author} {\bibinfo {author} {\bibfnamefont {A.~D.}\ \bibnamefont
  {Raki\'{c}}}, \bibinfo {author} {\bibfnamefont {A.~B.}\ \bibnamefont
  {Djuri\v{s}i\'{c}}}, \bibinfo {author} {\bibfnamefont {J.~M.}\ \bibnamefont
  {Elazar}},\ and\ \bibinfo {author} {\bibfnamefont {M.~L.}\ \bibnamefont
  {Majewski}},\ }\bibfield  {title} {\bibinfo {title} {Optical properties of
  metallic films for vertical-cavity optoelectronic devices},\ }\href
  {https://doi.org/10.1364/AO.37.005271} {\bibfield  {journal} {\bibinfo
  {journal} {Appl. Optics}\ }\textbf {\bibinfo {volume} {37}},\ \bibinfo
  {pages} {5271} (\bibinfo {year} {1998})}\BibitemShut {NoStop}%
\bibitem [{\citenamefont {Feist}\ \emph {et~al.}(2017)\citenamefont {Feist},
  \citenamefont {Bach}, \citenamefont {{Rubiano da Silva}}, \citenamefont
  {Danz}, \citenamefont {Möller}, \citenamefont {Priebe}, \citenamefont
  {Domröse}, \citenamefont {Gatzmann}, \citenamefont {Rost}, \citenamefont
  {Schauss}, \citenamefont {Strauch}, \citenamefont {Bormann}, \citenamefont
  {Sivis}, \citenamefont {Schäfer},\ and\ \citenamefont {Ropers}}]{Feist2016}%
  \BibitemOpen
  \bibfield  {author} {\bibinfo {author} {\bibfnamefont {A.}~\bibnamefont
  {Feist}}, \bibinfo {author} {\bibfnamefont {N.}~\bibnamefont {Bach}},
  \bibinfo {author} {\bibfnamefont {N.}~\bibnamefont {{Rubiano da Silva}}},
  \bibinfo {author} {\bibfnamefont {T.}~\bibnamefont {Danz}}, \bibinfo {author}
  {\bibfnamefont {M.}~\bibnamefont {Möller}}, \bibinfo {author} {\bibfnamefont
  {K.~E.}\ \bibnamefont {Priebe}}, \bibinfo {author} {\bibfnamefont
  {T.}~\bibnamefont {Domröse}}, \bibinfo {author} {\bibfnamefont {J.~G.}\
  \bibnamefont {Gatzmann}}, \bibinfo {author} {\bibfnamefont {S.}~\bibnamefont
  {Rost}}, \bibinfo {author} {\bibfnamefont {J.}~\bibnamefont {Schauss}},
  \bibinfo {author} {\bibfnamefont {S.}~\bibnamefont {Strauch}}, \bibinfo
  {author} {\bibfnamefont {R.}~\bibnamefont {Bormann}}, \bibinfo {author}
  {\bibfnamefont {M.}~\bibnamefont {Sivis}}, \bibinfo {author} {\bibfnamefont
  {S.}~\bibnamefont {Schäfer}},\ and\ \bibinfo {author} {\bibfnamefont
  {C.}~\bibnamefont {Ropers}},\ }\bibfield  {title} {\bibinfo {title}
  {Ultrafast transmission electron microscopy using a laser-driven field
  emitter: Femtosecond resolution with a high coherence electron beam},\ }\href
  {https://doi.org/https://doi.org/10.1016/j.ultramic.2016.12.005} {\bibfield
  {journal} {\bibinfo  {journal} {Ultramicroscopy}\ }\textbf {\bibinfo {volume}
  {176}},\ \bibinfo {pages} {63} (\bibinfo {year} {2017})}\BibitemShut
  {NoStop}%
\bibitem [{\citenamefont {Yannai}\ \emph {et~al.}(2023)\citenamefont {Yannai},
  \citenamefont {Adiv}, \citenamefont {Dahan}, \citenamefont {Wang},
  \citenamefont {Gorlach}, \citenamefont {Rivera}, \citenamefont {Fishman},
  \citenamefont {Kr\"uger},\ and\ \citenamefont {Kaminer}}]{Yannai2023}%
  \BibitemOpen
  \bibfield  {author} {\bibinfo {author} {\bibfnamefont {M.}~\bibnamefont
  {Yannai}}, \bibinfo {author} {\bibfnamefont {Y.}~\bibnamefont {Adiv}},
  \bibinfo {author} {\bibfnamefont {R.}~\bibnamefont {Dahan}}, \bibinfo
  {author} {\bibfnamefont {K.}~\bibnamefont {Wang}}, \bibinfo {author}
  {\bibfnamefont {A.}~\bibnamefont {Gorlach}}, \bibinfo {author} {\bibfnamefont
  {N.}~\bibnamefont {Rivera}}, \bibinfo {author} {\bibfnamefont
  {T.}~\bibnamefont {Fishman}}, \bibinfo {author} {\bibfnamefont
  {M.}~\bibnamefont {Kr\"uger}},\ and\ \bibinfo {author} {\bibfnamefont
  {I.}~\bibnamefont {Kaminer}},\ }\bibfield  {title} {\bibinfo {title}
  {Lossless monochromator in an ultrafast electron microscope using near-field
  thz radiation},\ }\href {https://doi.org/10.1103/PhysRevLett.131.145002}
  {\bibfield  {journal} {\bibinfo  {journal} {Phys. Rev. Lett.}\ }\textbf
  {\bibinfo {volume} {131}},\ \bibinfo {pages} {145002} (\bibinfo {year}
  {2023})}\BibitemShut {NoStop}%
\bibitem [{\citenamefont {Kuzyk}\ \emph {et~al.}(2012)\citenamefont {Kuzyk},
  \citenamefont {Schreiber}, \citenamefont {Fan}, \citenamefont {Pardatscher},
  \citenamefont {Roller}, \citenamefont {Högele}, \citenamefont {Simmel},
  \citenamefont {Govorov},\ and\ \citenamefont {Liedl}}]{kuzyk2012}%
  \BibitemOpen
  \bibfield  {author} {\bibinfo {author} {\bibfnamefont {A.}~\bibnamefont
  {Kuzyk}}, \bibinfo {author} {\bibfnamefont {R.}~\bibnamefont {Schreiber}},
  \bibinfo {author} {\bibfnamefont {Z.}~\bibnamefont {Fan}}, \bibinfo {author}
  {\bibfnamefont {G.}~\bibnamefont {Pardatscher}}, \bibinfo {author}
  {\bibfnamefont {E.-M.}\ \bibnamefont {Roller}}, \bibinfo {author}
  {\bibfnamefont {A.}~\bibnamefont {Högele}}, \bibinfo {author} {\bibfnamefont
  {F.}~\bibnamefont {Simmel}}, \bibinfo {author} {\bibfnamefont
  {A.}~\bibnamefont {Govorov}},\ and\ \bibinfo {author} {\bibfnamefont
  {T.}~\bibnamefont {Liedl}},\ }\bibfield  {title} {\bibinfo {title} {Dna-based
  self-assembly of chiral plasmonic nanostructures with tailored optical
  response},\ }\href {https://doi.org/10.1038/nature10889} {\bibfield
  {journal} {\bibinfo  {journal} {Nature}\ }\textbf {\bibinfo {volume} {483}},\
  \bibinfo {pages} {311} (\bibinfo {year} {2012})}\BibitemShut {NoStop}%
\bibitem [{\citenamefont {Lan}\ \emph {et~al.}(2015)\citenamefont {Lan},
  \citenamefont {Lu}, \citenamefont {Shen}, \citenamefont {Ke}, \citenamefont
  {Ni},\ and\ \citenamefont {Wang}}]{lan_au_2015}%
  \BibitemOpen
  \bibfield  {author} {\bibinfo {author} {\bibfnamefont {X.}~\bibnamefont
  {Lan}}, \bibinfo {author} {\bibfnamefont {X.}~\bibnamefont {Lu}}, \bibinfo
  {author} {\bibfnamefont {C.}~\bibnamefont {Shen}}, \bibinfo {author}
  {\bibfnamefont {Y.}~\bibnamefont {Ke}}, \bibinfo {author} {\bibfnamefont
  {W.}~\bibnamefont {Ni}},\ and\ \bibinfo {author} {\bibfnamefont
  {Q.}~\bibnamefont {Wang}},\ }\bibfield  {title} {\bibinfo {title} {Au nanorod
  helical superstructures with designed chirality},\ }\href
  {https://doi.org/10.1021/ja511333q} {\bibfield  {journal} {\bibinfo
  {journal} {J. Am. Chem. Soc.}\ }\textbf {\bibinfo {volume} {137}},\ \bibinfo
  {pages} {457} (\bibinfo {year} {2015})}\BibitemShut {NoStop}%
\bibitem [{\citenamefont {Gaida}\ \emph {et~al.}(2023)\citenamefont {Gaida},
  \citenamefont {Lourenço-Martins}, \citenamefont {Yalunin}, \citenamefont
  {Feist}, \citenamefont {Sivis}, \citenamefont {Hohage}, \citenamefont
  {Garcia~de Abajo},\ and\ \citenamefont {Ropers}}]{gaida2023}%
  \BibitemOpen
  \bibfield  {author} {\bibinfo {author} {\bibfnamefont {J.}~\bibnamefont
  {Gaida}}, \bibinfo {author} {\bibfnamefont {H.}~\bibnamefont
  {Lourenço-Martins}}, \bibinfo {author} {\bibfnamefont {S.}~\bibnamefont
  {Yalunin}}, \bibinfo {author} {\bibfnamefont {A.}~\bibnamefont {Feist}},
  \bibinfo {author} {\bibfnamefont {M.}~\bibnamefont {Sivis}}, \bibinfo
  {author} {\bibfnamefont {T.}~\bibnamefont {Hohage}}, \bibinfo {author}
  {\bibfnamefont {J.}~\bibnamefont {Garcia~de Abajo}},\ and\ \bibinfo {author}
  {\bibfnamefont {C.}~\bibnamefont {Ropers}},\ }\bibfield  {title} {\bibinfo
  {title} {Lorentz microscopy of optical fields},\ }\href
  {https://doi.org/10.1038/s41467-023-42054-3} {\bibfield  {journal} {\bibinfo
  {journal} {Nat. Commun.}\ }\textbf {\bibinfo {volume} {14}},\ \bibinfo
  {pages} {6545} (\bibinfo {year} {2023})}\BibitemShut {NoStop}%
\bibitem [{\citenamefont {Gaida}\ \emph {et~al.}(2024)\citenamefont {Gaida},
  \citenamefont {Lourenço-Martins}, \citenamefont {Sivis}, \citenamefont
  {Rittmann}, \citenamefont {Feist}, \citenamefont {Garcia~de Abajo},\ and\
  \citenamefont {Ropers}}]{gaida2023xiv}%
  \BibitemOpen
  \bibfield  {author} {\bibinfo {author} {\bibfnamefont {J.~H.}\ \bibnamefont
  {Gaida}}, \bibinfo {author} {\bibfnamefont {H.}~\bibnamefont
  {Lourenço-Martins}}, \bibinfo {author} {\bibfnamefont {M.}~\bibnamefont
  {Sivis}}, \bibinfo {author} {\bibfnamefont {T.}~\bibnamefont {Rittmann}},
  \bibinfo {author} {\bibfnamefont {A.}~\bibnamefont {Feist}}, \bibinfo
  {author} {\bibfnamefont {J.}~\bibnamefont {Garcia~de Abajo}},\ and\ \bibinfo
  {author} {\bibfnamefont {C.}~\bibnamefont {Ropers}},\ }\href
  {https://doi.org/10.1038/s41566-024-01380-8} {\bibinfo {title} {Attosecond
  electron microscopy by free-electron homodyne detection}} (\bibinfo {year}
  {2024})\BibitemShut {NoStop}%
\bibitem [{\citenamefont {Nabben}\ \emph {et~al.}(2023)\citenamefont {Nabben},
  \citenamefont {Kuttruff}, \citenamefont {Stolz}, \citenamefont {Ryabov},\
  and\ \citenamefont {Baum}}]{nab2023}%
  \BibitemOpen
  \bibfield  {author} {\bibinfo {author} {\bibfnamefont {D.}~\bibnamefont
  {Nabben}}, \bibinfo {author} {\bibfnamefont {J.}~\bibnamefont {Kuttruff}},
  \bibinfo {author} {\bibfnamefont {L.}~\bibnamefont {Stolz}}, \bibinfo
  {author} {\bibfnamefont {A.}~\bibnamefont {Ryabov}},\ and\ \bibinfo {author}
  {\bibfnamefont {P.}~\bibnamefont {Baum}},\ }\bibfield  {title} {\bibinfo
  {title} {Attosecond electron microscopy of sub-cycle optical dynamics},\
  }\href {https://doi.org/10.1038/s41586-023-06074-9} {\bibfield  {journal}
  {\bibinfo  {journal} {Nature}\ }\textbf {\bibinfo {volume} {619}},\ \bibinfo
  {pages} {63} (\bibinfo {year} {2023})}\BibitemShut {NoStop}%
\bibitem [{\citenamefont {Fang}\ \emph {et~al.}(2024)\citenamefont {Fang},
  \citenamefont {Kuttruff}, \citenamefont {Nabben},\ and\ \citenamefont
  {Baum}}]{fang2024}%
  \BibitemOpen
  \bibfield  {author} {\bibinfo {author} {\bibfnamefont {Y.}~\bibnamefont
  {Fang}}, \bibinfo {author} {\bibfnamefont {J.}~\bibnamefont {Kuttruff}},
  \bibinfo {author} {\bibfnamefont {D.}~\bibnamefont {Nabben}},\ and\ \bibinfo
  {author} {\bibfnamefont {P.}~\bibnamefont {Baum}},\ }\bibfield  {title}
  {\bibinfo {title} {Structured electrons with chiral mass and charge},\ }\href
  {https://doi.org/10.1126/science.adp9143} {\bibfield  {journal} {\bibinfo
  {journal} {Science}\ }\textbf {\bibinfo {volume} {385}},\ \bibinfo {pages}
  {183} (\bibinfo {year} {2024})},\ \Eprint
  {https://arxiv.org/abs/https://www.science.org/doi/pdf/10.1126/science.adp9143}
  {https://www.science.org/doi/pdf/10.1126/science.adp9143} \BibitemShut
  {NoStop}%
\end{thebibliography}
\end{document}